\documentclass[pra,twocolumn,superscriptaddress,amsmath,amssymb]{revtex4}
\usepackage{graphicx}
\usepackage{dsfont}

\newcommand{\ket}[1]{| #1 \rangle}
\newcommand{\braket}[2]{\langle #1 | #2 \rangle}
\newcommand{\ketbra}[2]{| #1 \rangle \langle #2 |}

\begin{document}
\title{Verifying single-mode nonclassicality beyond negativity in phase space}
\author{Jiyong Park}
\email{jiyong.park@hanbat.ac.kr}
\affiliation{School of Basic Sciences, Hanbat National University, Daejeon 34158, Korea}
\author{Jaehak Lee}
\affiliation{School of Computational Sciences, Korea Institute for Advanced Study, Seoul 02455, Korea}
\author{Hyunchul Nha}
\email{hyunchul.nha@qatar.tamu.edu}
\affiliation{Department of Physics, Texas A\&M University at Qatar, Education City, P.O.Box 23874, Doha, Qatar}
\date{\today}

\begin{abstract}
While negativity in phase space is a well-known signature of nonclassicality, a wide variety of nonclassical states require their characterization beyond negativity. We establish a framework of nonclassicality in phase space that addresses nonclassical states comprehensively with a direct experimental evidence. This includes the negativity of phase-space distribution as a special case and further analyzes quantum states with positive distributions effectively. We prove that it detects all nonclassical Gaussian states and all non-Gaussian states of arbitrary dimension remarkably by examining three phase-space points only. Our formalism also provides an experimentally accessible lower bound for a nonclassicality measure based on trace distance. Importantly, this foundational approach can be further adapted to constitute practical tests in two directions looking into particle and wave nature of bosonic systems, via an array of nonideal on-off detectors and coarse-grained homodyne measurement, respectively. All these tests are practically powerful in characterizing nonclasssical states reliably against noise, making a versatile tool for a broad range of quantum systems in quantum technologies.
\end{abstract}

\maketitle

\section{Introduction}
Describing a quantum state of light or matter in phase space, e.g., Wigner function \cite{Wigner1932}, is profoundly important to study quantum dynamics. It is a crucial tool to delineate the boundary between classical and quantum physics widely used in quantum optics \cite{Barnett}, continuous variable (CV) quantum informatics \cite{Braunstein2005, Weedbrook2012} and other fields of quantum science \cite{Weinbub2018}. As a classical phase-space distribution takes non-negative values like a probability distribution, a negativity emerging in quantum distribution is regarded as a signature of nonclassicality. However, negativity is just one aspect of multifaceted nonclassicality characterizing only a subset of nonclassical states. There exist quantum states with positive distributions that can nevertheless be classified as nonclassical, e.g., a squeezed state of light that is a key resource for CV quantum informatics \cite{Weedbrook2012} and single photons under a high-loss channel  that are elementary information carriers for quantum informatics \cite{Milburn, Kok}. 

Nonclassical states are essential resources broadly for quantum informatics generating entangled states \cite{Kim2002, Wang2002, Asboth2005, Nha2008, Tahira2009}, providing advantage for quantum metrology \cite{Yadin2018, Kwon2019,Tan2019} and quantum computation \cite{Bartlett2015, Ralph2015}, etc.. A recent resource theory identified all quantum non-Gaussian states, even with positive Wigner functions, as a resource for quantum tasks, e.g., subchannel discrimination \cite{Takagi}, which was further generalized to all CV nonclassical states \cite{Regula}. It is thus crucial to establish a framework that can widely analyze nonclassical states beyond negativity. Specific properties were often used to characterize nonclassical states such as squeezing and photon-number statistics (sub-Poissonian) \cite{Agarwal, Klyshko,Simon1,Simon2} extended also to multi-mode cases \cite{Lee, Simon3}. Distillation of nonclassicality can also be used to verify the nonclassicality of an initial state, however, requiring multiple copies of the same nonclasscial state and postselection \cite{Hage2007, Filip2013, Filip2014}. More broadly, a quantum state tomography may be used to obtain complete information on a state thereby confirming nonclassicality \cite{Lvovsky2009}. However, it requires extensive measurements for sufficient data, and more seriously, an optimization process to find a physical state closest to obtained data. The data itself does not directly represent a legitimate quantum state rendering its significance weaker. It is necessary to characterize nonclassicality by examining phase-space in a faithful and resource-efficient way. 

Adhering to negativity as a nonclassical feature, some works proposed to display negativity by modifying phase-space distributions, e.g., a regularized $P$-function under filtering process \cite{Kiesel1,Kiesel2}. Other distributions closely related to the so-called {\it s}-parametrized functions \cite{Barnett} were also studied in view of photon statistics from on-off detectors \cite{Luis2015,Bohmann2018}. Phase-space inequalities were also obtained by combining different $s$-parametrized functions useful to some extent \cite{Bohmann20}. Nevertheless, it is worth asking if the original Wigner function contains substantial information on nonclassicality beyond negativity. In this respect, Banaszek and W{\'o}dkiewicz proposed a Bell test examining four phase-space points to manifest nonlocality of two-mode states with positive Wigner functions \cite{Banaszek1999}, which were extended to generalized quasiprobability distributions \cite{SWLee2009} and genuine multipartite nonlocality \cite{SWLee2013, Adesso2014, Xu2017}. The works in \cite{Park2015a, Park2015b} demonstrated Bell-like tests also for single-mode nonclassicality and quantum non-Gaussianity. While conceptually remarkable and practically useful, these methods do not address a broad range of nonclassical states, e.g., squeezed states with purity $<0.86$ are out of reach. 

In this article, we propose a hierarchy of nonclassicality criteria in phase space that yields an efficient and broadly applicable test for CV systems. 
Our formalism addresses the Wigner function at $\frac{n(n+1)}{2}$ phase-space points progressively ($n=1,2,\dots$). It includes the negativity of Wigner function at $n = 1$. Remarkably, it can detect all nonclassical Gaussian states and all non-Gaussian states of arbitrary dimension at the next level $n=2$, i.e., looking into three phase-space points only. This opens a new possibility for a faithful and efficient test. We show that our foundational approach can constitute two practical tests characterizing nonclassical states reliably and efficiently from a particle and a wave point of views, respectively. It thus makes our method a versatile tool for a wide range of CV systems in quantum physics and technologies. We illustrate the practical power of our approach by examples. Our proposed approach is fruitful also in other aspects. It provides an experimentally accessible lower bound for nonclassical distance defined via trace norm \cite{Hillery1987}, which is hard to obtain even theoretically. It can also be further extended to identify quantum non-Gaussianity \cite{Filip2011, Jezek2011, Straka2014, Straka2018, Lachman2018, Genoni2013, Hughes2014, Kuhn2018, Happ2018, Takagi2018, Albarelli2018, Park2019a, Park2019b, Lee2019, Park2017} under energy constraint. 

\section{Criteria}
Let us start with a general condition on classicality. A classical state, i.e., a mixture of coherent states, must satisfy 
\begin{equation} \label{eq:CS}
		\int d^{2} \alpha P_{\rho_c} ( \alpha ) | f ( \alpha ) |^{2} \geq 0,
	\end{equation}
for an arbitrary $f ( \alpha )$ since its Sudarshan-Glauber-$P$ function $P_{\rho_c} ( \alpha )$ is positive definite \cite{Sudarshan,Glauber}.
Our aim is to establish criteria that deal with the Wigner function at discrete phase-space points by choosing $f ( \alpha )$ properly. Not only providing a fundamental insight, the Wigner-function approach also leads to two general practical tests broadly applicable for CV systems, as shown later.

To our aim, invoking the convolution between the $P$-function and the $s$-parametrized function  \cite{Barnett}
	\begin{equation}\label{eq:conv}
		W_{\rho} ( \alpha; s ) = \frac{2}{\pi (1-s)} \int d^{2} \beta P_{\rho} ( \beta ) e^{- \frac{2 | \beta - \alpha |^{2}}{1-s}},
	\end{equation}
we choose $f ( \alpha ) = \sum_{i=1}^{n} c_{i} e^{- |\alpha-\beta_{i}|^{2}}$ ($s = 0$ for Wigner function, $c_i$, $\beta_i$: arbitrary complex numbers). 
It yields $\int d^{2} \alpha P_{\rho} ( \alpha ) | f ( \alpha ) |^{2} = \sum_{i,j=1}^{n} c_{i}^{*} c_{j} \mathcal{M}_{ij}^{(n)} \geq 0,$
 where 
\begin{eqnarray}
\mathcal{M}_{ij}^{(n)} = \frac{\pi}{2} W_{\rho} \bigg( \frac{\beta_{i}+\beta_{j}}{2} \bigg) e^{ - \frac{1}{2} |\beta_{i}-\beta_{j}|^{2}}. 
\end{eqnarray}
For the classicality to hold for arbitrary $c_i$'s, we deduce the following theorem.

{\it Theorem}. An $n\times n$ matrix $\mathcal{M}^{(n)}$ with its elements given by Eq. (3) must be positive semidefinite for a classical state, i.e., $\mathcal{M}^{(n)} \succeq \mathbf{0}$ for all $n \in [ 1, \infty )$, with arbitrary $\{ \beta_{1}, \beta_{2}, ... \beta_{n} \}$. In other words, we verify nonclassicality if there exists a nonpositive $\mathcal{M}^{(n)} \nsucceq \mathbf{0}$ for any $n$.

\section{Hierarchy}
By its construction, $\mathcal{M}^{(n+1)} \succeq \mathbf{0}$ implies $\mathcal{M}^{(n)} \succeq \mathbf{0}$ since the matrix $\mathcal{M}^{(n+1)}$ includes $\mathcal{M}^{(n)}$ as its submatrix. 
That is, there naturally occurs a hierarchy of criteria with $n$ increasing. If nonclassicality is confirmed at the level of $n$, it must be so at the next levels of $n+1$, etc., but the converse is not always true.

Our formulation includes the negativity of Wigner function at the lowest $n=1$, $\mathcal{M}^{(1)} =\frac{\pi}{2}W(\beta)\nsucceq \mathbf{0}$. Then, it is fundamentally interesting, and practically important, to know how many phase-space points are required to verify noclassicality for states with positive Wigner functions. We prove below that our method can detect nonclassical states comprehensively using only three points $\{ \beta_1,\frac{\beta_{1}+\beta_{2}}{2},\beta_2 \}$ on a line, i.e., 
$\mathcal{M}^{(2)} \nsucceq \mathbf{0}$ with 
\begin{equation}
\mathcal{M}^{(2)}=\frac{\pi}{2}\begin{pmatrix} 
W( \beta_1) & W( \frac{\beta_{1}+\beta_{2}}{2}) e^{ - \frac{1}{2} |\beta_{1}-\beta_{2}|^{2}}\\W( \frac{\beta_{1}+\beta_{2}}{2}) e^{ - \frac{1}{2} |\beta_{1}-\beta_{2}|^{2}} & W( \beta_2)
\end{pmatrix}.
\end{equation}

\section{Geometric interpretation}
Before demonstrating its usefulness, let us briefly discuss the meaning of the classicality condition $\mathcal{M}^{(2)} \ge \mathbf{0}$.
One readily finds that all coherent states satisfy $W_{\rm coh}( \frac{\beta_{1}+\beta_{2}}{2}) e^{ - \frac{1}{2} |\beta_{1}-\beta_{2}|^{2}}=\sqrt{W_{\rm coh}( \beta_1)W_{\rm coh}( \beta_2)}$ yielding $\mathcal{M}^{(2)} \ge \mathbf{0}$ 
for arbitrary $\{\beta_1,\beta_2\}$.
The linearity of $\mathcal{M}^{(n)}$ with respect to states, $\mathcal{M}_{\sum p_i\rho_i}^{(2)}=\sum p_i\mathcal{M}_{\rho_i}^{(2)}$, then makes a general classicality condition $\mathcal{M}^{(2)} \ge \mathbf{0}$ for a mixture of coherent states. For a classical state, we thus see that the Wigner function at midpoint $\frac{\beta_{1}+\beta_{2}}{2}$ must be bounded by the geometric mean of the Wigner functions at end points $\beta_1$ and $\beta_2$, importantly with a scaling factor $e^{ - \frac{1}{2} |\beta_{1}-\beta_{2}|^{2}}$. In fact, this factor results from the commutator $[\hat{a},\hat{a}^\dag]=1$ representing the size of vacuum fluctuation.

	\begin{figure}[!tb]
		\includegraphics[scale=0.5]{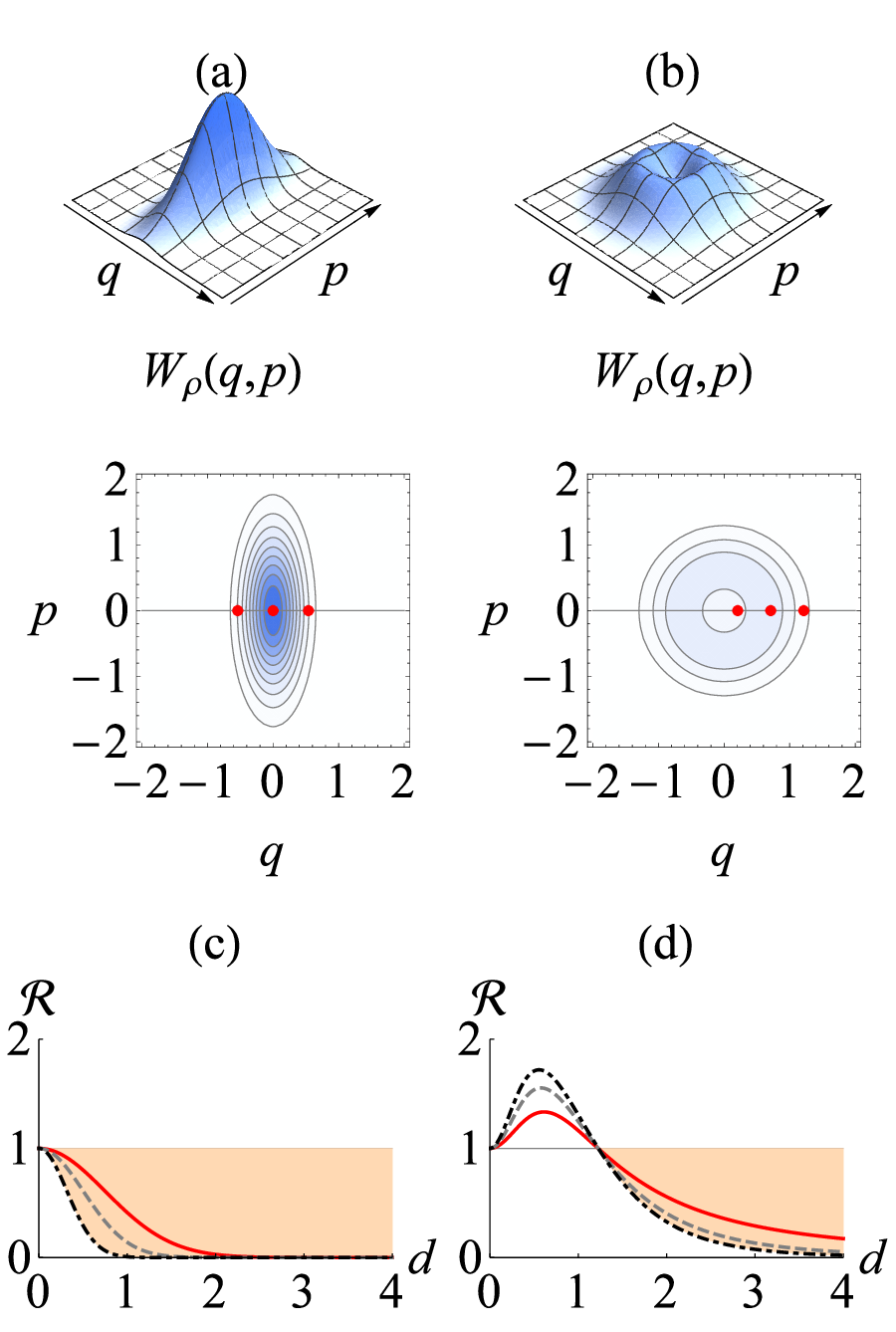}
		\caption{Wigner function of (a) squeezed vacuum ($r=0.5$) and (b) a lossy single photon $0.6 \ketbra{0}{0} + 0.4 \ketbra{1}{1}$. Red dots in contour plot represent the phase-space points for $\mathcal{M}^{(2)}$ test. (c) $\mathcal{R} = \frac{W_{\rho} (-d ) W_{\rho} ( d )}{W_{\rho} ( 0 )^{2} \exp ( - 4d^{2} )}$ for $x$-squeezed states with squeezing $r - r_{c} = 0.1$ (red solid) $r - r_{c} = 0.2$ (gray dashed) $r - r_{c} = 0.4$ (black dot-dashed) and (d)  $\mathcal{R} =\frac{W_{\rho} ( 0 ) W_{\rho} ( 2d )}{W_{\rho} ( d )^{2} \exp ( - 4d^{2} )}$ for a Fock state $|n\rangle$ under 80 \% loss channel for $n=1$ (red solid), $n=2$ (gray dashed), $n=3$ (black dot-dashed). $\mathcal{R}<1$ (shaded region) confirms nonclassicality for a wide range of displacement $d$.}
		\label{fig:W3P}
	\end{figure}

{\it Gaussian states}. 
Every single-mode Gaussian state can be expressed as a displaced squeezed thermal state 
\begin{equation}
		\sigma = \hat{D} ( \gamma ) \hat{S} ( r, \phi ) \rho_{th} ( \bar{n} ) \hat{S}^{\dag} ( r, \phi ) \hat{D}^{\dag} ( \gamma).
	\end{equation}
Here $\hat{S} ( r, \phi ) = \exp [ - \frac{r}{2} \left( e^{2i \phi} ( \hat{a}^{\dag} )^{2} - e^{- 2i \phi} \hat{a}^{2} \right) ]$ is a squeezing operator with strength $r$ and angle $\phi$ of squeezing axis. $\rho_{th} ( \bar{n} )=\sum_{n = 0}^{\infty} \frac{\bar{n}^{n}}{( \bar{n} + 1 )^{n + 1}} \ketbra{n}{n}$ is a thermal state with mean number $\bar{n}$. We can readily show $\mathcal{M}^{(2)} \nsucceq \mathbf{0}$ by taking three points along a squeezed axis [Fig. 1(a)], with two end points at a distance $2d$ and the middle point at the origin. Our test turns out to be successful for a wide range of $d$ as shown in Fig. 1(c). Without loss of generality, we consider an $x$-squeezed thermal state ($\gamma,\phi=0$), whose Wigner function is given by $W_{\sigma} ( q ,p ) = \frac{2\mu}{\pi} e^{- 2e^{2(r-r_{c})}q^{2}} e^{- 2e^{-2(r+r_{c})}p^{2}}$, with purity $\mu=(1+2\bar{n})^{-1}$ and critical squeezing $r_{c} = - \frac{1}{2} \log \mu$. Section S1 of the Supplemental Material (SM) \cite{Supple} gives its lowest eigenvalue of $\mathcal{M}^{(2)}$ as 
	\begin{equation}\label{eq:MEVG}
		\lambda_{\min, \sigma} = - 2 \mu e^{-(r-r_{c})\coth(r-r_{c})} \sinh (r-r_{c}) <0,
	\end{equation}
confirming nonclassicality for every squeezed state $r>r_c$, pure or mixed. 

{\it Non-Gaussian states}: More importantly, the three-points test $\mathcal{M}^{(2)} \nsucceq \mathbf{0}$ can detect a broad range of non-Gaussian states. We first demonstrate its success for all non-Gaussian states of arbitrary truncation in Fock space. This includes as examples all noisy Fock states having positive Wigner functions. In Sec. S3 of SM \cite{Supple}, we further demonstrate that it can be extended to states of practical relevance having infinite Fock-state components.

The Wigner function of an arbitrary Fock-space truncated state (FSTS), $\rho = \sum_{i,j=0}^{N} \rho_{jk} \ketbra{j}{k}$, takes a form 
$W_{\rho} ( \alpha ) = \sum_{i,j=0}^{N} \rho_{jk} W_{\ketbra{j}{k}} ( \alpha )$, with $W_{\ketbra{j}{k}} ( \alpha )$ given in Sec. S2 of SM \cite{Supple}.
As the case of negative Wigner functions is already treated at $n=1$, we focus on the case of positive Wigner functions. 
Choosing $\beta_{1} = 2d e^{i \varphi}$ and $\beta_{2} = 0$ gives $\det \mathcal{M}^{(2)} = \frac{\pi^{2}}{4} \left[W_{\rho} ( 2de^{i \varphi} ) W_{\rho} ( 0 ) - W_{\rho}^{2} ( d e^{i \varphi} ) e^{-4d^{2}}\right]. $
We thus look into $\mathcal{R} ( d ) = \frac{W_{\rho} ( 2de^{i \varphi} ) W_{\rho} ( 0 )}{W_{\rho}^{2} ( de^{i \varphi} ) e^{-4d^{2}}}$ whose value less than 1 verifies nonclassicality. $\mathcal{R} ( d )$ is a continuous function of $d$ satisfying $\mathcal{R} ( 0 ) = 1$. For the FSTS, we always find $\lim_{d \rightarrow \infty} \mathcal{R} ( d ) = 0$ with details in Sec. S2 of SM \cite{Supple}.
Therefore, there must be a finite $d$ satisfying $\mathcal{R}(d) < 1$ confirming nonclassicality. Remarkably, it works regardless of $\varphi$, i.e. insensitive to the axis of three points.

As an illustration, in Fig. 1, we plot the ratio $\mathcal{R}$ for (c) squeezed states and (d) Fock states under a 80\%-loss channel. We confirm nonclassicality, $\mathcal{R}<1$, for a broad range of displacement $d$.

\section{Nonclassicality distance}
It is also a topic of great interest to quantify the degree of nonclassicality for a given state $\rho$. 
A typical approach is to measure a distance between $\rho$ and its closest classical state $\rho_c$ as ${\mathcal N}_d(\rho)\equiv\frac{1}{2} \min_{\rho_c \in \mathcal{C}} || \rho - \rho_c ||_{1}$, with $||\cdot||_{1}$ the trace norm and $\mathcal{C}$ the set of classical states. This is, however, very hard to obtain even if the state is completely known. Our formalism provides a lower bound for this nonclassical distance \cite{Hillery1987, Nair2017} enabling its practical estimation. 
With details in Sec. S4 of SM \cite{Supple}, we obtain
	\begin{equation} \label{eq:est}
		{\mathcal N}_d(\rho) \geq - \frac{\lambda_{\min}}{2n},
	\end{equation}
where $\lambda_{\min}$ is the least eigenvalue of $\mathcal{M}^{(n)}$ at the level $n$.

At $n=1$, Eq.~(\ref{eq:est}) shows that a negative value in phase space directly provides a reliable estimate for nonclassical distance.
We can further estimate the nonclassical distance of a state with a postive Wigner function by using $\mathcal{M}^{(2)}$. For instance, for a general Gaussian state $\sigma$,
	\begin{equation}
		{\mathcal N}_d(\sigma) \geq \frac{\mu}{2} e^{-(r-r_{c})\coth(r-r_{c})} \sinh (r-r_{c}),
	\end{equation}
using Eq. (6), which is beyond the results in Refs. \cite{Hillery1987, Nair2017} addressing only pure Gaussian states. We also establish connection between our approach and nonclassical depth \cite{Lee1991} in S9 of SM \cite{Supple}.

\section{QNG}
Furthermore, our formalism also leads to a criterion on quantum non-Gaussianity (QNG) manifesting that a state cannot be a mixture of Gaussian states. 
\cite{Park2015a,Filip2011,Jezek2011,Straka2014,Straka2018,Lachman2018,Genoni2013,Hughes2014,Happ2018,Park2019a,Lee2019}. QNG has recently attracted much attention in CV quantum informatics as there exist numerous quantum tasks essentially requiring it beyond Gaussian resources, e.g., quantum computation \cite{Lloyd,Menicucci}, entanglement distillation \cite{Eisert,Fiurasek,Giedke}, and error correction \cite{Niset}.
 
\begin{figure}[!tb]
		\includegraphics[scale=0.6]{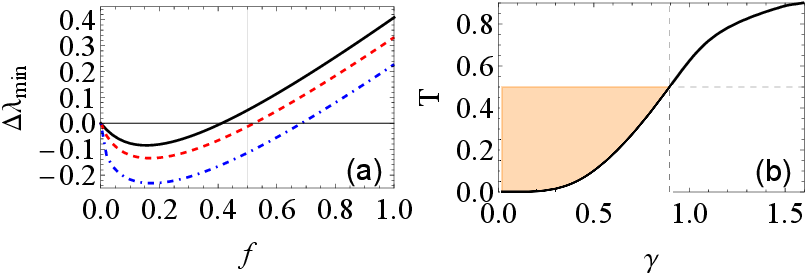}
		\caption{(a) $\Delta \lambda_{\min} = \mathcal{B} ( E ) - \lambda_{\min}$ for the state $\hat{S}(r) \{ f \ketbra{2}{2} + (1-f) \ketbra{0}{0} \} \hat{S}^{\dag} (r)$ with $r=0$ (blue dot-dashed), $r=0.2$ (red dashed) and $r=0.5$ (black solid). The QNG-detectable region, $\Delta \lambda_{\min} > 0$, broadens with squeezing $r$. (b) The QNG of a four-component cat state $\ket{\rm C}\sim\ket{\gamma}+\ket{\gamma e^{i\frac{\pi}{2}}}+\ket{\gamma e^{i\pi}}+\ket{\gamma e^{i\frac{3\pi}{2}}}$ \cite{cat state} is confirmed under a loss channel ($T$: transmittance) for each $\gamma$. Black solid represents the minimum $T$ above which QNG is verified by Eq. (9). The yellow region represents the case of positive Wigner function. }
		\label{fig:S02m}
	\end{figure}

With details in Sec. S5 of SM \cite{Supple}, if the least eigenvalue of $\mathcal{M}_{\rho}^{(2)}$ for a state $\rho$ with energy $E$ satisfies 
\begin{equation}
		\lambda_{\min} < \mathcal{B} ( E ) \equiv - \frac{2 \sqrt{E}}{(\sqrt{E+1}+\sqrt{E})^{\sqrt{1+E^{-1}}}}.
	\end{equation}
 it confirms QNG. 
In Fig.~\ref{fig:S02m}, we plot $\Delta \lambda_{\min} = \mathcal{B} ( E ) - \lambda_{\min}$ for a non-Gaussian state $\rho = \hat{S}(r) \{ f \ketbra{2}{2} + (1-f) \ketbra{0}{0} \} \hat{S}^{\dag} (r)$ with squeezing $\hat{S} ( r ) = e^{\frac{r}{2}(a^\dag)^2-\frac{r}{2}a^2}$.
As seen from Fig.~\ref{fig:S02m}(a), our criterion detects QNG with $f < \frac{1}{2}$ (positive Wigner function) for a squeezing $r \gtrsim 0.237$. 
Note that the squeezing operation does not create QNG as it is a Gaussian operation. In this context, the result also represents the QNG of $f \ketbra{2}{2} + (1-f) \ketbra{0}{0}$ without squeezing. 
A recent ion-trap experiment realized a measurement in squeezed Fock basis, $\{\hat{S} ( r )|n\rangle: n=0,1,\cdots\}$
\cite{Ion}. This can be adopted to verify QNG of states $\hat{S} ( r )\rho_{\rm nG} \hat{S}^\dag ( r )$ without performing squeezing on a non-Gaussian state $\rho_{\rm nG}$ enhancing the range of QNG detection. 
Fig.~\ref{fig:S02m}(b) gives another example of a positive Wigner function with its QNG verified.

\section{Practical tests}
The Wigner function corresponds to the number parity after displacement, i.e., $W_{\rho} ( \alpha ) = \frac{2}{\pi} \mathrm{tr} [ \hat{D}^{\dag} ( \alpha )\rho \hat{D} ( \alpha ) (-1)^{\hat{n}}  ] $. It is routinley measured in various systems, e.g., ion-trap \cite{Wineland,Park2015a} and circuit-QED \cite{Sun}, for which our proposed test $\mathcal{M}^{(2)}$ can thus directly characterize nonclassicality. 
On the other hand, we can also derive alternative, practical, schemes out of Wigner-function framework, which can test nonclassicality reliably and efficiently against experimental imperfections.  
First, we present a generalized formalism to use on-off detectors registering photons without photon-number resolving (PNR). 
Second, we also present a marginal version of Wigner-function test, i.e., using $M(q)=\int dp W_{\rho} (q,p)$, which can be measured by homodyne detection well established for a wide variety of quantum systems including quantum optics \cite{Lvovsky2009}, trapped ion \cite{Gerritsma}, atomic ensemble \cite{Fernholz}, circuit cavity QED \cite{Mallet,Eichler}, and optomechanics \cite{Hertzberg,Vanner}. 
Both of our proposed tests are powerful against noise with wide applicability.

	\begin{figure}[!tb]
          \includegraphics[scale=0.4]{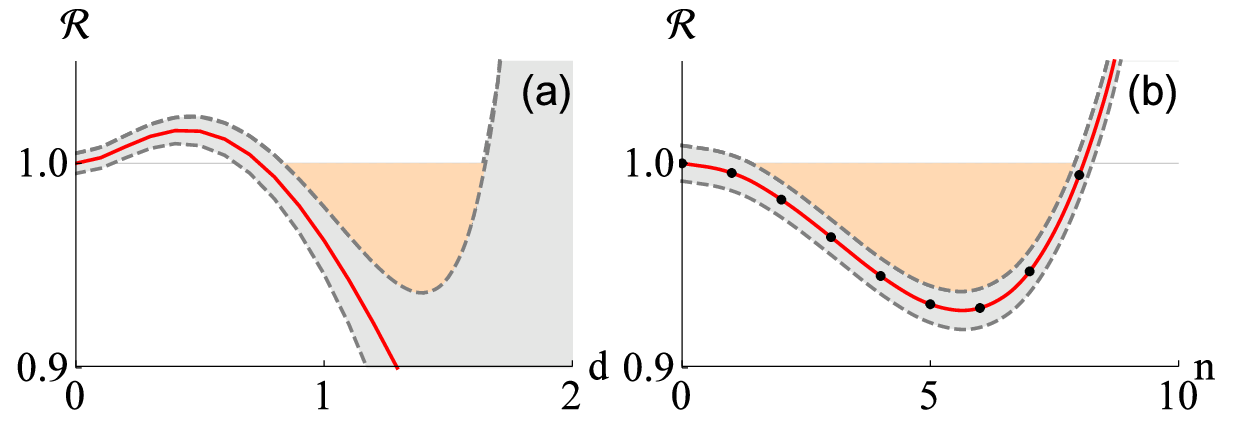}
		\caption{(a) Testing a Fock state $|n=2\rangle$ under 50\% loss channel mixed with a thermal photon $\bar n=0.05$ using  $N=1$ on-off detectors. $\mathcal{R} = \frac{W_{\rho} (d_{1}; s) W_{\rho} (d_{1}+2d; s)}{W_{\rho} (d_{1}+d; s)^{2} \exp ( - 4d^{2} )}$ ($s=-1.86$, red solid) against displacement $d$ with $d_{1}=1$. 
(b) Homodyne test for a phase-diffused squeezed state with $\mathcal{R} = \frac{M_{\rho}^{\sigma} [ -n ] M_{\rho}^{\sigma} [ n ]}{M_{\rho}^{\sigma} [0]^{2} \exp ( - 4n^{2}\sigma^{2} )}$ (red solid) against $n$ with binning size $\sigma = 0.1$. $n$: bin number for quadrature $q=n\sigma$. Grey shades represent the size of error due to finite data (a) $\sim10^5$ and (b) $\sim10^6$. }
		\label{fig:RatioCGL1}
	\end{figure}

{\it On-off detector array}. When an input light is equally divided via beam-splitting to impinge on $N$ on-off detectors, the probability of $k$-detectors clicking is \cite{Sperling2012a}
	\begin{equation} \label{eq:PCS}
		p_{k} [ \rho ] = \mathrm{tr} \left[ \rho : \frac{N}{k!(N-k)!} ( e^{-\frac{\eta \hat{n}}{N}} )^{N-k} ( 1 - e^{-\frac{\eta \hat{n}}{N}})^{k} : \right],
	\end{equation}
with $\eta$ detector efficiency and $:\hat {O}:$ normal-ordering. 

The counting statistics $p_k$ above can also be obtained via the time-multiplexing approach using a single detector \cite{Bohmann2018}.  
We first generalize our criterion to $s$-parametrized function $W_{\rho} ( \alpha; s )$ \cite{Barnett} to use the counting statistics $p_{k} [ \rho ]$ from on-off detectors.
Choosing $f ( \alpha ) = \sum_{i=1}^{n} c_{i} e^{ - \frac{| \alpha - \beta_{i} |^{2}}{1-s} }$ in Eq. (1) with the convolution in Eq. (2),  
we obtain $\int d^{2} \alpha P_{\rho} ( \alpha ) | f ( \alpha ) |^{2} = \sum_{i,j=1}^{n} c_{i}^{*} c_{j} \mathcal{M}_{ij}^{(s,n)} \geq 0,$
where
	\begin{align} \label{eq:coeff_s}
		\mathcal{M}_{ij}^{(s)} = \frac{\pi (1-s)}{2} W_{\rho} \bigg( \frac{\beta_{i}+\beta_{j}}{2} ; s \bigg) e^{- \frac{1}{1-s} \frac{| \beta_{i} - \beta_{j} |^{2}}{2}}.
	\end{align}
Our classicality condition is readily generalized to $\mathcal{M}^{(s,n)} \succeq \mathbf{0}$ for an arbitrary $s$ using elements in Eq. \eqref{eq:coeff_s}.

We find the connection between the $s$-parametrized functions and the counting statistics $p_k$ in S6 of SM \cite{Supple} as
	\begin{equation} \label{eq:sampling}
		\begin{pmatrix} W_{\rho} ( \alpha; s_{0} ) \\ W_{\rho} ( \alpha; s_{1} ) \\ \cdots \\ W_{\rho} ( \alpha; s_{N-1} ) \\ 1 \end{pmatrix} = T^{-1} \begin{pmatrix} p_{0} \\ p_{1} \\ \cdots \\ p_{N-1} \\ p_{N} \end{pmatrix},
	\end{equation}
with each $s_{m} = 1 - \frac{2N}{(N-m)\eta}$ ($m=0,\cdots,N-1$).
Eq. \eqref{eq:sampling} means that $N$ different $s$-parametrized distributions $W_{\rho} ( \alpha; s_{m} )$ are determined by the counting statistics $\{p_0,\cdots,p_N\}$ obtained for a displaced state $\hat{D}^{\dag} ( \alpha )\rho \hat{D} ( \alpha )$. 
Furthermore, we prove in Sec. S2 of SM \cite{Supple} that $\mathcal{M}^{(s,n=2)}$ (three points test) can detect all nonclassical Gaussian and non-Gaussian states (FSTSs), importantly for an {\it arbitrary} $s$. 
This broader applicability beyond Wigner function makes our test robust against noise. 

Let us illustrate the case of testing a Fock state $|n\rangle$ under 50\% loss channel mixed with a thermal photon $\bar n=0.05$ by using only $N=1$ on-off detector of efficiency $\eta=0.7$ \cite{Nam}. We further consider an error due to finite data acquisition $\sim10^5$ (Sec. S8 of SM \cite{Supple}). Our 3-points test adopting $W_{\rho} ( \alpha; s=-1.86)$ is accomplished with $s_m = 1 - \frac{2N}{(N-m)\eta}$ for $m=0$. As shown in Fig. 3(a), there exists a range of displacement to detect nonclassicality substantially beating the error. For instance, we have the signal to noise ratio as $\frac{1-\mathcal{R}}{\Delta \mathcal{R}}=2.48$ at $d=1.1$. 
We also demonstrate the successful detection for other noisy Fock states with error analysis in Sec. S8 of SM \cite{Supple}. 

{\it Homodyne test}. We next present a test using a marginal distribution $M(q)=\int dp W_{\rho} (q,p)$. Homodyne detection to measure $M(q)$ is highly efficient, but requires a careful analysis. It is because that the actual homodyne data is coarse-grained due to finite binning, which may lead to a false detection of nonclassical effects \cite{Schneeloch,Tasca,Park2014}. Let $\sigma$ be the binning size of homodyne data. Then all data in the range  $[-\sigma/2,\sigma/2]$ belong to the same bin yielding a coarse-grained distribution $M_{\rho}^{\sigma} [ n ] \equiv\int_{-\sigma/2}^{\sigma/2} d \delta M_{\rho} ( n \sigma + \delta )$ ($n$: bin number representing mean quadrature $q=n\sigma$). A classicality condition $\mathcal{M}^{(H)}\ge0$ then emerges with its elements  
\begin{equation}
\mathcal{M}_{ij}^{(H)}\equiv\frac{\pi}{2} M_{\rho}^{\sigma} [ m_{i} + m_{j} + k ] e^{- 2 (m_{i}-m_{j})^{2} \sigma^{2}},
\end{equation}
where $k$ can be either 0 or 1, with details in Sec. S7 of SM \cite{Supple}.

We prove in SM \cite{Supple} that this marginal test even with a coarse-grained information detects all nonclassical Gaussian and non-Gaussian states (FSTSs). In Fig. 3(b), we show the result for a squeezed state ($r=0.3$) under phase-diffusion, $\mathcal{D} [ \rho ] = \int d \phi \sqrt{\frac{1}{2 \pi \Delta^{2}}} e^{- \frac{\phi^{2}}{2 \Delta^{2}}} e^{i \hat{n} \phi} \rho e^{-i \hat{n} \phi}$ leaving no squeezing at $\Delta=1.2$. Our homodyne test under coarse-graining ($\sigma=0.1$) clearly manifests nonclassicality over 7 standard deviation, $\frac{1-\mathcal{R}}{\Delta \mathcal{R}}=7.11$. We also illustrate other cases in Sec. S8 of SM \cite{Supple}.

\section{Conclusion}
A phase-space approach usually provides us with a valuable insight into quantum physics \cite{Kolobov}. While negativity is one manifestation of nonclassicality, recent studies made it clear that all nonclassical states even without negativity are valuable resources for quantum information science \cite{Kim2002, Wang2002, Asboth2005, Nha2008,Takagi}. It is thus critically important to establish a comprehensive framework of addressing nonclassical states with and without negativity covering a wide range of quantum systems. We have introduced a hierarchy of nonclassicality conditions that can address nonclassicality beyond negativity effectively and efficiently. Our approach makes it possible to analyze all nonclassical Gaussian states and non-Gaussian states using three phase-space points. Our formalism further provides a lower bound for nonclassical distance and a criterion to detect quantum non-Gaussianity with positive Wigner functions. 
Remarkably, our foundational approach also constitutes two practical tests looking into particle nature (number parity) and wave nature (marginal distribution), making a versatile tool for CV systems broadly. 
We illustrated the practical power of our tests adopting nonideal on-off detectors without resolving photon numbers and coarse-grained homodyne detection, respectively.  

We hope our work could further stimulate works related to nonclassical effects from both a fundamental and a practical perspective. Our approach here clearly indicates that the information on nonclassicality is sufficiently imbedded in phase space even at a few points. Our geometric interpretation on classicality has stipulated the relation among the values of Wigner function, which is fundamentally associated with quantum fluctuation represented by a commutation relation or uncertainty principle. This seems worthwhile to further purse in studying nonclassicality for quantum multipartite systems as well. In the near term, we anticipate our framework can be useful for both theoretical and experimental analysis of quantum systems. In particular, our proposed tests can address all different CV systems including quantum optics, nano- or opto-mechanics, atomic ensemble, and circuit cavity QED, and so on.

{\it Note added}. We recently became aware of a closely related work by Bohmann, Agudelo and Sperling \cite{Bohmann2020}. We note that our main idea and some results were earlier presented at an international conference \cite{ICSSUR2017}.

\section*{ACKNOWLEDGMENTS}
J.P. acknowledges support by the National Research Foundation of Korea (NRF) grant funded by the Korea government (MSIT) (NRF-2019R1G1A1002337). J.L. is supported by a KIAS Individual Grant (CG073101) at Korea Institute for Advanced Study. H.N. is supported by a grant NPRP13S-0205-200258 from Qatar National Research Fund.

\bibliographystyle{apsrev}

\setcounter{equation}{0}
\renewcommand{\theequation}{S\arabic{equation}}
\setcounter{figure}{0}
\renewcommand{\thefigure}{S\arabic{figure}}

\section*{S1. Optimal phase-space test for a Gaussian state}
We first consider a $2 \times 2$ matrix $A$ whose matrix elements are given by
	\begin{align}
		A_{ij} = &  F \bigg( \frac{x_{i}+x_{j}}{2}, \frac{y_{i}+y_{j}}{2} \bigg) \nonumber \\
		& \times \exp \bigg[ - \frac{c}{4} (x_{1}-x_{2})^{2} - \frac{c}{4} (y_{1}-y_{2})^{2} \bigg],
	\end{align}
where $F ( x, y ) = e^{- ax^{2} - by^{2}}$ is a Gaussian function with $a > c > b > 0$. 
Then, we show that the minimum lowest eigenvalue of the matrix $A$ is given by
	\begin{equation} \label{eq:MLEV}
		\lambda_{\min} =  - \bigg( 1 - \frac{c}{a} \bigg) \bigg( \frac{c}{a} \bigg)^{\dfrac{c}{a-c}}.
	\end{equation}	

For a given function $F ( x, y ) = e^{- ax^{2} - by^{2}}$, we can show that the points $(x_{1},y_{1})$ and $(x_{2},y_{2})$ minimizing the lowest eigenvalue of $A$ must be (1) on $x$-axis, i.e., $y_{1} = y_{2} = 0$ and (2) symmetric with respect to the origin, i.e., $x_{1}+x_{2}=0$. First, the lowest eigenvalue of $A$ is given by
	\begin{align} \label{eq:LEV1}
		\lambda & = \frac{A_{11}+A_{22}}{2} - \sqrt{\Big(\frac{A_{11}-A_{22}}{2}\Big)^{2}+A_{12}^{2}} \nonumber \\
		& = \frac{A_{11}+A_{22}}{2} - \sqrt{\Big(\frac{A_{11}-A_{22}}{2}\Big)^{2}+A_{11}A_{22}R},
	\end{align}
where
	\begin{equation}
		R = \exp \bigg[ \frac{a-c}{2} (x_{1}-x_{2})^{2} + \frac{b-c}{2} (y_{1}-y_{2})^{2} \bigg].
	\end{equation}
For fixed values of $A_{11} = u$ and $A_{22} = v$, the set of the points $(x_{1},y_{1})$ and $(x_{2},y_{2})$ satisfying $F ( x_{1}, y_{1} ) = u$ and $F ( x_{2}, y_{2} ) = v$ form two ellipses having the same center, directrix and major axis. 
Under the conditions $a-c>0$ and $b-c<0$, the ratio $R$ is maximized at
	\begin{align} \label{eq:OP1}
		( x_{1}, y_{1} ) & = \bigg( \pm \sqrt{- \frac{\log u}{a}}, 0 \bigg), \nonumber \\
		( x_{2}, y_{2} ) & = \bigg( \mp \sqrt{- \frac{\log v}{a}}, 0 \bigg).
	\end{align}

We now set $y_{1} = y_{2} = 0$ and rewrite Eq. \eqref{eq:LEV1} as
	\begin{align} \label{eq:LEV2}
		\lambda = & \frac{A_{11}+A_{22}}{2} \nonumber \\
		& - \sqrt{\Big(\frac{A_{11}+A_{22}}{2}\Big)^{2} + A_{11} A_{22} (R-1)},
	\end{align}
Now that $\lambda$ in Eq. \eqref{eq:LEV2} is a function of two variables $A_{11} = u$ and $A_{22} = v$, we further fix its product $A_{11}A_{22} = uv\equiv w$. 
This also fixes the value of $x_{1}^2+x_{2}^2=-\frac{1}{a}\log{w}$. Note that $X - \sqrt{X^{2}+Y}$ with $Y > 0$ decreases as $X$ decreases or $Y$ increases. Using the relation between the arithmetic mean and the geometric mean,  we observe that $A_{11}+A_{22}$ and $A_{11}A_{22}(R-1)$ are minimized and maximized, respectively, when $|x_{1}|=|x_{2}|$ and $u=v$ for a given $uv=w$.

All things considered together with $|x_{1}|=|x_{2}|=x$, we now need to optimize
	\begin{equation} \label{eq:LEV3}
		\lambda = \exp ( - a x^{2} ) -  \exp ( - c x^{2} ).
	\end{equation}
Examining its first derivative, we obtain the minimum lowest eigenvalue of $A$ in Eq. \eqref{eq:MLEV} at
	\begin{equation} \label{eq:recipe}
		( x_{1}, x_{2} ) =  \bigg( \pm \sqrt{\frac{\log \frac{a}{c}}{a-c}}, \mp \sqrt{\frac{\log \frac{a}{c}}{a-c}} \bigg).
	\end{equation}

The Wigner function of  the $x$-squeezed thermal state is given by $W_{\sigma} ( \alpha = q + ip ) = \frac{2\mu}{\pi} e^{- 2e^{2(r-r_{c})}q^{2}} e^{- 2e^{-2(r+r_{c})}p^{2}}$,
with purity $\mu=(1+2\bar{n})^{-1}$ and critical squeezing $r_{c} = - \frac{1}{2} \log \mu$. 
In view of Eq. (S1), its parameters read $a=2e^{2(r-r_{c})}$, $b=2e^{-2(r+r_{c})}$ and $c=2$ satisfying the condition $a > c > b > 0$ for a nonclassical state $r>r_c$. 
Therefore, its least eigenvalue of $\mathcal{M}^{(2)}$ becomes
	\begin{equation}
		\lambda_{\min, \sigma} = - 2 \mu e^{-(r-r_{c})\coth(r-r_{c})} \sinh (r-r_{c}) <0.
	\end{equation}
It occurs at the optimal choice of phase-space points $\{ \beta_{1}, \beta_{2} \} = \{\pm \sqrt{\frac{r-r_{c}}{e^{2(r-r_{c})}-1}}, \mp \sqrt{\frac{r-r_{c}}{e^{2(r-r_{c})}-1}}\}$ from the recipe in Eq. \eqref{eq:recipe}.

\section*{S2. Power of criteria using $s$-parametrized functions}
We here show that all nonclassical Gaussian states and non-Gaussian states of arbitrary Fock-space truncation can be detected via the matrix $\mathcal{M}^{(s,2)}$ criterion for any $s$. This naturally includes the case of Wigner function test, $s=0$.

{\it Gaussian states}---
The $s$-parametrized quasiprobability function of a $x$-squeezed thermal state is given by
	\begin{equation}
		W_{\sigma}^{(s)} ( \alpha = q + ip ) = \frac{\sqrt{ab}}{\pi} \exp ( - a q^{2} - b p^{2} ),
	\end{equation}
where
	\begin{align}
		a & = \frac{2}{e^{-2(r-r_{c})} - s}, \nonumber \\
		b & = \frac{2}{e^{2(r+r_{c})} - s},
	\end{align}
with $r$ and $r_{c}$ representing the squeezing strength and the critical squeezing strength for nonclassicality, respectively. 
Using the minimum lowest eigenvalue of $\mathcal{M}_{\sigma}^{(s,2)}$ in Eq. \eqref{eq:MLEV} of S1 
	\begin{equation}
		\lambda_{\min} = - \frac{(1-s) \sqrt{ab}}{2} \bigg( 1 - \frac{c}{a} \bigg) \bigg( \frac{c}{a} \bigg)^{\dfrac{c}{a-c}} < 0,
	\end{equation}
with the parameters $a$ and $b$ for the Gaussian states, we see that it becomes negative for all nonclassical states $r > r_{c}$ regardless of $s<1$.\\
{\it Non-Gaussian states}---
The $s$-parametrized quasiprobability function of an arbitrary Fock-space truncated state (FSTS), $\rho = \sum_{j=0}^{N} \sum_{k=0}^{N} \rho_{jk} \ketbra{j}{k}$, is written as
	\begin{equation}
		W_{\rho} ( \alpha; s ) = \sum_{j=0}^{N} \sum_{k=0}^{N} \rho_{jk} W_{\ketbra{j}{k}} ( \alpha; s ),
	\end{equation}
where $W_{\ketbra{j}{k}} ( \alpha; s )$ for $j \geq k$ \cite{SI-Park2015b} is given by
	\begin{align}
		W_{\ketbra{j}{k}} ( \alpha; s ) = & \frac{2}{\pi (1-s)} \exp \bigg( - \frac{2 |\alpha|^{2}}{1-s} \bigg) \bigg( \frac{s+1}{s-1} \bigg)^{k} \nonumber \\
		& \times \sqrt{\frac{k!}{j!}} \bigg( \frac{2 \alpha}{1-s} \bigg)^{j-k} L_{k}^{(j-k)} \bigg( \frac{4 |\alpha|^{2}}{1-s^{2}} \bigg),
	\end{align}
with a generalized Laguerre polynomial $L_{n}^{(m)} (z) = \sum_{\ell=0}^{n} \frac{(n+m)!}{(n-\ell)!(m+\ell)!\ell!} (-z)^{\ell}$ of degree $n$ and $W_{\ketbra{j}{k}} ( \alpha,s ) = W_{\ketbra{k}{j}} ( \alpha^{*},s )$ for $j < k$. 
For the matrix $\mathcal{M}^{(s,2)}$, if we set $\beta_{1} = 2 d e^{i \varphi}$ and $\beta_{2} = 0$, its determinant becomes
	\begin{align}
		\det \mathcal{M}^{(s,2)} = & \frac{\pi^{2}(1-s)^{2}}{4} \bigg\{ W_{\rho} ( 2de^{i \varphi}; s ) W_{\rho} ( 0; s ) \nonumber \\
& - W_{\rho} ( de^{i \varphi}; s )^{2} \exp \bigg( - \frac{4d^{2}}{1-s} \bigg) \bigg\}.
	\end{align}
We thus investigate a function of $d$
\begin{equation}
\mathcal{R} ( d ) = \frac{W_{\rho} ( 2de^{i \varphi},s ) W_{\rho} ( 0,s )}{W_{\rho}^{2} ( de^{i \varphi},s ) e^{-\frac{4d^{2}}{1-s}}}
\end{equation}
 whose value less than 1 confirms nonclassicality. $\mathcal{R} ( d )$ is a continuous function of $d$ satisfying $\mathcal{R} ( 0 ) = 1$. 
For an FSTS, we always find $\lim_{d \rightarrow \infty} \mathcal{R} ( d ) = 0$ with a dominant contribution given by $W_{\ketbra{N}{N}}$ terms, as $W_{\rho} ( 2d e^{i \varphi} ) \propto e^{-\frac{8d^{2}}{1-s}} d^{2N}$ and $W_{\rho} ( d e^{i \varphi} )^{2} \propto e^{-\frac{4d^{2}}{1-s}} d^{4N}$ for $d \gg 1$. Therefore, there must be a finite $d$ satisfying $\mathcal{R}(d) < 1$, i.e. $\det \mathcal{M}^{(2)} < 0$ for all $s<1$.

\begin{figure}[ht]
          \includegraphics[scale=0.6]{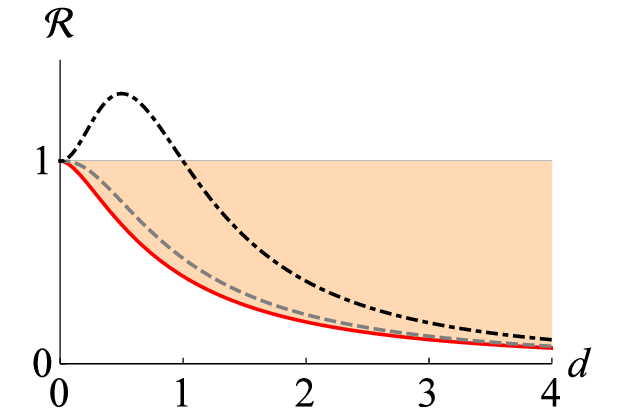}
		\caption{Testing a superposition state $\frac{1}{\sqrt{2}}\left(|0\rangle+|1\rangle\right)$ under a 50\% loss channel using the Wigner function test.  $\mathcal{R} =\frac{W_{\rho} ( 0 ) W_{\rho} ( 2d )}{W_{\rho} ( d )^{2} \exp ( - 4d^{2} )}$, where $d$ represents the distance from the origin along $x$-axis (red solid), $\frac{1}{\sqrt{2}}(x+p)$-axis (gray dashed) and $p$-axis (black dot-dashed), respectively. $\mathcal{R}<1$ confirms the detection of nonclassicality for a wide range of $d$ in each case.}
		\label{fig:L01}
	\end{figure}

Remarkably, the above proof works regardless of $\varphi$, i.e. insensitive to the axis of three points. We illustrate it by an example of non-rotationally symmetric state in phase space, that is, a superposition state $\frac{1}{\sqrt{2}}\left(|0\rangle+|1\rangle\right)$ under a 50 \% loss. In Fig. \ref{fig:L01}, we show the results of testing it by choosing three points along $x$-axis (red solid), $\frac{1}{\sqrt{2}}(x+p)$-axis (gray dashed) and $p$-axis (black dot-dashed), respectively. We can clearly see that the test is successful for a wide range of displacement $d$ whatever axis is taken.

\section*{S3. Non-Gaussian states with infinite Fock-state components}

We here illustrate the detection of non-Gaussian states having infinite Fock-state components that are of current practical interest for CV quantum informatics. 

\subsection{Dephased cat states}
The Wigner function of dephased cat state $\rho = \mathcal{N} ( \ketbra{\gamma}{\gamma} + \ketbra{-\gamma}{-\gamma} + f \ketbra{\gamma}{-\gamma} + f \ketbra{-\gamma}{\gamma} )$ with the normalization factor $\mathcal{N} = [ 2 + 2 f \exp ( -2\gamma^{2} ) ]^{-1}$ is given by
	\begin{align}
		W_{\rho} ( q+ip ) = & \frac{2 \mathcal{N}}{\pi} [ e^{-2(q-\gamma)^{2}-2p^{2}} + e^{-2(q+\gamma)^{2}-2p^{2}} \nonumber \\
		& + 2f e^{-2q^{2}-2p^{2}} \cos (4 \gamma p) ].
	\end{align}
It covers both of the even-cat states and the odd-cat states by taking $f>0$ and $f<0$, respectively. 
The dephased cat state has a positve Wigner function if and only if $|f| < e^{-2 \gamma^{2}}$, which can be shown by using arithmetic-geometric mean inequality. That is, 
	\begin{align}
		W_{\rho} ( q+ip ) & \geq \frac{2 \mathcal{N}}{\pi} [ 2 e^{-2q^{2}-2\gamma^{2}-2p^{2}} + 2f e^{-2q^{2}-2p^{2}} \cos (4 \gamma p) ] \nonumber \\
		& = \frac{4 \mathcal{N}}{\pi} e^{-2q^{2}-2p^{2}} [ e^{-2\gamma^{2}} + f \cos (4 \gamma p) ] \nonumber \\
		& \geq \frac{4 \mathcal{N}}{\pi} e^{-2q^{2}-2p^{2}} [ e^{-2\gamma^{2}} - |f| ],
	\end{align}
which indicates that the Wigner function becomes positive if $|f| < e^{-2\gamma^{2}}$. In addition, $W_{\rho} ( 0 ) = \frac{4\mathcal{N}}{\pi} ( e^{-2\gamma^{2}} + f )$ and $W_{\rho} ( \frac{i \pi}{4 \gamma} ) = \frac{4\mathcal{N}}{\pi} e^{-\frac{\pi^{2}}{8 \gamma^{2}}} ( e^{-2\gamma^{2}} - f )$, so the Wigner function cannot be positive when $|f| \geq e^{-2\gamma^{2}}$.

Setting $\beta_{1} = 0$ and $\beta_{2} = \frac{i \pi}{4 \gamma}$, we observe that
	\begin{equation}
		\mathcal{R} = \frac{W_{\rho} ( \beta_{1} ) W_{\rho} ( \beta_{2} )}{W_{\rho} ( \frac{\beta_{1}+\beta_{2}}{2} )^{2} e^{-|\beta_{1}-\beta_{2}|^{2}}} = 1 - f^{2} e^{4 \gamma^{2}},
	\end{equation}
which manifests that our method can detect all nonclassical dephased cat states regardless of $\gamma$ and $f$.

\subsection{Photon added coherent states}
Another infinite-dimensional quantum state of current interest is the photon-added coherent state, $|\Psi\rangle\sim a^\dag|\gamma\rangle$. 
As $|\Psi\rangle$ is a pure non-Gaussian state with negativity, let us treat a realistic situation under a loss channel, i.e. $\rho={\rm Tr}_E\left[U_{BS} |\Psi\rangle\langle\Psi|\otimes|0\rangle\langle0|_EU_{BS}^\dag\right]$, where $U_{BS}$ represents a beam splitting operation. We first note that  $a^\dag|\gamma\rangle=a^\dag {\hat D}(\gamma)|0\rangle={\hat D}(\gamma)(a^\dag+\gamma^*)|0\rangle={\hat D}(\gamma)(|1\rangle+\gamma^*|0\rangle)$. That is, $|\Psi\rangle$ is nothing but a displaced FSTS. 
As the local displacement followed by the beam splitting can be expressed in different order as the beam splitting followed by another displacement, we finally see that the mixed state $\rho$ is just a lossy FSTS followed by a displacement. We have already proved that any FSTS can be detected under our formalism. As the final displacement can be incorporated to the choice of three phase-space points accordingly, it proves that all photon added states under a loss channel can be detected as well. The same idea is readily extended also to the multiple-times photon-added coherent states, $|\Psi\rangle\sim a^{\dag m}|\gamma\rangle={\hat D}(\gamma)(a^\dag+\gamma^*)^m|0\rangle$, under a noisy channel.

\section*{S4. Nonclassicality distance}

The nonclassical distance is defined as ${\mathcal N}_d(\rho)\equiv\frac{1}{2} \min_{\rho_c \in \mathcal{C}} || \rho - \rho_c ||_{1}$, with $||\cdot||_{1}$ the trace norm and $\mathcal{C}$ the set of classical states. 
If ${\mathcal N}_d(\rho)=D$, it allows a decomposition $\rho = \rho_{c} + D ( \rho_{+} - \rho_{-} )$ where $\rho_{c}$ is its nearest classical state under the trace measure. $D \rho_{+}$ and $- D \rho_{-}$ represent the mixtures of eigenstates for $\rho - \rho_{c}$ with positive and negative eigenvalues, respectively ($\mathrm{tr} \rho_{+} = \mathrm{tr} \rho_{-} = 1$). 
For a general operator $\hat{O}$, we have
	\begin{equation}
		\mathrm{tr} [ \rho \hat{O} ] \geq \mathrm{tr} [ \rho_{c} \hat{O} ] + D ( \min_{\rho_{+} \in \mathcal{Q}} \mathrm{tr} [ \rho_{+} \hat{O} ] - \max_{\rho_{-} \in \mathcal{Q}} \mathrm{tr} [ \rho_{-} \hat{O} ] ),
	\end{equation}
where $\mathcal{Q}$ represents the set of all quantum states. 
We set $\hat{O} =\sum_{i,j=1}^{n} u_{i}^{*} u_{j} \hat{M}_{ij}$ with its connection to $\mathcal{M}_{ij}$ of Eq. (3) in main text as $\mathcal{M}_{ij} = \mathrm{tr} [ \rho \hat{M}_{ij} ]$. Here  $u = \{ u_{1}, u_{2}, ..., u_{n} \}$ is specifically taken to be the eigenvector for the lowest eigenvalue of $\mathcal{M}$. From $\mathrm{tr} [ \rho_{c} \hat{O} ] \geq 0$, we obtain
	\begin{equation}
		{\mathcal N}_d(\rho)=D \geq - \frac{\lambda_{\min}}{2n}.
	\end{equation}
We have above used $| \mathrm{tr} [ \sigma \hat{O} ] | \leq n$ for any state $\sigma \in \mathcal{Q}$ from $|\mathrm{tr} [ \sigma \hat{M}_{ij} ]| \leq 1$ and $\sum_{i=1}^{n} |u_{i}| \leq \sqrt{n}$.

\section*{S5. Quantum non-Gaussianity test}

Our formalism can further identify quantum non-Gaussianity (QNG), i.e. those states beyond a mixture of Gaussian states.  
Eq.~(S9) is the lowest eigenvalue of $\mathcal{M}^{(2)}$ for a Gaussian state, which can be rewritten in terms of energy $E = \mathrm{tr} [ \rho \hat{n} ]$. We verify QNG if the least eigenvalue of a given state is smaller than those of Gaussian states with the same $E$
	\begin{equation}
		\lambda_{\min} < \mathcal{B} ( E ) \equiv - \frac{2 \sqrt{E}}{(\sqrt{E+1}+\sqrt{E})^{\sqrt{1+E^{-1}}}}.
	\end{equation}

We here derive the above Gaussian bound $\mathcal{B} ( E )$ in the following steps. 
To begin with, note that the phase-space matrix $\mathcal{M}^{(n)}$ is linear with respect to states, i.e. $\mathcal{M}_{\sum p_i\rho_i}^{(2)}=\sum p_i\mathcal{M}_{\rho_i}^{(2)}$. 

(i) Suppose that a given state $\rho$ is a mixture of pure Gaussian states $\sigma_i$,  $\rho=\sum_ip_i\sigma_i$, with its energy $E={\rm tr}\{\rho \hat{n}\}=\sum_ip_i{\rm tr}\{\sigma_i  \hat{n}\}=\sum_ip_i E_i^G$. 
In this case, the smallest eigenvalue $\lambda$ of $\mathcal{M}_\rho^{(2)}$ cannot be less than the weighted sum of the smallest eigenvalues $\lambda_i$ of $\mathcal{M}_{\sigma_i}^{(2)}$, i.e. $\lambda\ge\sum_ip_i\lambda_i$.
This can be readily seen by considering the eigenvectors corresponding to the least eigenvalues as $|\lambda\rangle$ and $|\lambda_i\rangle$, 
i.e. $\langle\lambda|\mathcal{M}^{(2)}|\lambda\rangle=\lambda$ and $\langle\lambda_i|\mathcal{M}_{\sigma_i}^{(2)}|\lambda_i\rangle=\lambda_i$, respectively. 
Then, from $\mathcal{M}^{(2)}=\sum p_i\mathcal{M}_{\sigma_i}^{(2)}$, we obtain 
$\lambda=\langle\lambda|\mathcal{M}^{(2)}|\lambda\rangle=\sum p_i\langle\lambda|\mathcal{M}_{\sigma_i}^{(2)}|\lambda\rangle\ge\sum p_i\lambda_i$ due to the condition $\langle\lambda|\mathcal{M}_{\sigma_i}^{(2)}|\lambda\rangle\ge\langle\lambda_i|\mathcal{M}_{\sigma_i}^{(2)}|\lambda_i\rangle=\lambda_i$.

(ii) A pure Gaussian state is a displaced squeezed state, $\sigma = \hat{D} ( \gamma ) \hat{S} ( r, \phi ) |0\rangle\langle0| \hat{S}^{\dag} ( r, \phi ) \hat{D}^{\dag} ( \gamma)$, which has the energy $E=\sinh^2r+|\alpha|^2$. 
On the other hand, as we explained in the main text, the displacement does not affect the least eigenvalue of $\mathcal{M}_{\sigma}^{(2)}$, which is given by $\lambda_{\min, \sigma} = - 2 e^{-r\coth r} \sinh r$ in Eq. (S9). 
This means that the squeezed state without displacement is energy-efficient to achieve the same level of least eigenvalue. We thus consider only the squeezed states with energy $E=\sinh^2r$, which gives the expression 
	\begin{equation}
		\lambda_{\min, \sigma}^E=- \frac{2 \sqrt{E}}{(\sqrt{E+1}+\sqrt{E})^{\sqrt{1+E^{-1}}}}.
	\end{equation}
for its least eigenvalue. 

(iii) Importantly, we note that $\lambda_{\min, \sigma}^E$ above is a convex function of $E$ (Fig. 1). Thus, among all Gaussian states with the same average energy $E$, the pure squeezed state achieves the smallest eigenvalue. If a state is a mixture of pure squeezed states with $E=\sum_ip_iE_i$, its least eigenvalue cannot be less than that of a single pure squeezed states with energy $E$ due to the convexity of the function in Eq. (S21). Therefore, $\lambda_{\min, \sigma}^E$ represents the smallest possible eigenvalue of the matrix 
$\mathcal{M}_{\sigma}^{(2)}$ among all Gaussian states $\sigma$ with the energy $E$. 

Combining (i), (ii) and (iii), we obtain a QNG criterion. That is, if the least eigenvalue of $\mathcal{M}_{\rho}^{(2)}$ for a state $\rho$ with energy $E$ is less than $\mathcal{B} ( E )\equiv\lambda_{\min, \sigma}^E$, it confirms QNG. The state cannot be a mixture of Gaussian states.

	\begin{figure}
		\includegraphics[scale=0.6]{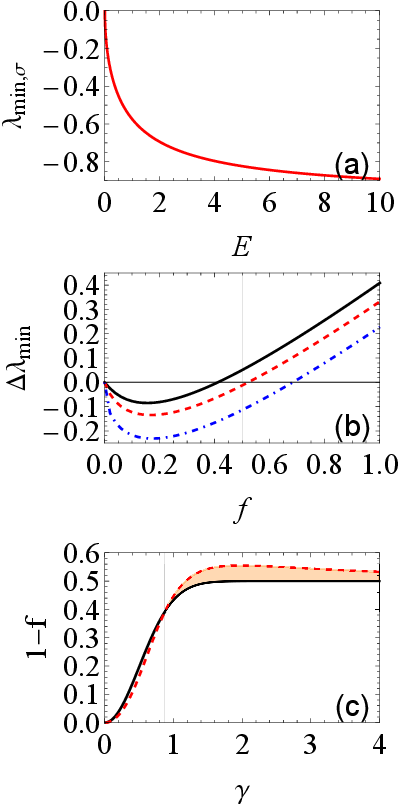}
		\caption{(a) $\lambda_{\min, \sigma}^E=- \frac{2 \sqrt{E}}{(\sqrt{E+1}+\sqrt{E})^{\sqrt{1+E^{-1}}}}$ as a function of $E$. We see that it is a convex function of $E$. (b) $\Delta \lambda_{\min} = \mathcal{B} ( E ) - \lambda_{\min}$ for $\hat{S}(r) \{ f \ketbra{2}{2} + (1-f) \ketbra{0}{0} \} \hat{S}^{\dag} (r)$ with $r=0$ (blue dot-dashed), $r=0.2$ (red dashed) and $r=0.5$ (black solid). The QNG-detectable region, $\Delta \lambda_{\min} > 0$, becomes larger with a higher squeezing. (c) Detection of QNG for the mixed cat state $ f \ketbra{\rm C}{\rm C} + (1-{\it f}) \ketbra{0}{0}$, with a cat state $\ket{\rm C}\sim\ket{\gamma}+\ket{-\gamma}$. Black solid represents the value of $1-f$ at which the Wigner function becomes positive, above which our criterion detects QNG for the mixed cat state (red dashed). }
		\label{fig:QNG-SI}
	\end{figure}

In Fig.~\ref{fig:QNG-SI}, we plot $\Delta \lambda_{\min} = \mathcal{B} ( E ) - \lambda_{\min}$ for a non-Gaussian state $\rho = \hat{S}(r) \{ f \ketbra{2}{2} + (1-f) \ketbra{0}{0} \} \hat{S}^{\dag} (r)$ with squeezing $\hat{S} ( r ) = e^{\frac{r}{2}(a^\dag)^2-\frac{r}{2}a^2}$. For a fraction $f > \frac{1}{2}$, the state has the negativity of Wigner function, which is a trivial evidence of QNG. 
As can be seen from Fig.~\ref{fig:QNG-SI}, our criterion can detect QNG with $f < \frac{1}{2}$ for a squeezing $r \gtrsim 0.237$. 
Note that the squeezing operation does not change QNG as it is a Gaussian operation. In this respect, the result in Fig.~\ref{fig:QNG-SI} also represents the QNG of $f \ketbra{2}{2} + (1-f) \ketbra{0}{0}$ without squeezing. 

A recent ion-trap experiment realized a measurement in squeezed Fock basis, $\{\hat{S} ( r )|n\rangle: n=0,1,\cdots\}$ by controlling the interaction between the spin and the motional states \cite{SI-Kienzler2016}. This measurement can be adopted to verify QNG of those states $\hat{S} ( r )\rho_{\rm nG} \hat{S}^\dag ( r )$ without performing squeezing on a non-Gaussian state $\rho_{\rm nG}$.

Fig. S2 (c) gives another example of a positive Wigner function with its QNG verified. 
The QNG of a mixed cat state $ f \ketbra{\rm C}{\rm C} + (1-{\it f}) \ketbra{0}{0}$, with a cat state $\ket{\rm C}\sim\ket{\gamma}+\ket{-\gamma}$, is confirmed up to the vacuum fraction $1-f$ for each $\gamma$. Black solid represents the value of $1-f$ at which the Wigner function becomes positive, above which our criterion detects QNG for the mixed cat state (red dashed). The yellow-filled region thus represents the detection of QNG for a non-Gaussian state with a positive Wigner function.

\section*{S6. Connection between the $s$-parametrized functions and the counting statistics from on-off detectors}
We start with the counting statistics Eq.~(10) in main text and a relation between the expectation value of a normally ordered operator for a quantum state $\rho$ with its Glauber-P function, i.e., $\mathrm{tr} [ \rho : f ( \hat{n} ) : ] = \int d^{2} \beta P_{\rho} ( \beta ) f ( |\beta|^{2} )$ for an arbitrary well-defined function $f$ \cite{SI-BarnettRadmore}.  We first obtain
	\begin{align}
		p_{k} [ \rho ] & = \mathrm{tr} [ \rho : \frac{N}{k!(N-k)!} ( e^{-\frac{\eta \hat{n}}{N}} )^{N-k} ( 1 - e^{-\frac{\eta \hat{n}}{N}})^{k} : ], \nonumber \\
		& = \int d^{2} \beta P_{\rho} ( \beta ) \binom{N}{k} ( e^{-\frac{\eta |\beta|^{2}}{N}} )^{N-k} ( 1 - e^{-\frac{\eta |\beta|^{2}}{N}})^{k}.
	\end{align}	
Using $P_{\hat{D}^{\dag} ( \alpha ) \rho \hat{D} ( \alpha )} ( \beta ) = P_{\rho} ( \alpha + \beta )$, we get
	\begin{align}
		& p_{k} [ \hat{D}^{\dag} ( \alpha ) \rho \hat{D} ( \alpha ) ] \nonumber \\
		& = \int d^{2} \beta P_{\rho} ( \beta + \alpha ) \binom{N}{k} ( e^{-\frac{\eta |\beta|^{2}}{N}} )^{N-k} ( 1 - e^{-\frac{\eta |\beta|^{2}}{N}})^{k} \nonumber \\
		& = \int d^{2} \beta P_{\rho} ( \beta + \alpha ) \binom{N}{k} \sum_{m=0}^{k} \binom{k}{m} (-1)^{k-m} e^{-\frac{(N-m)\eta}{N} |\beta|^{2}} \nonumber \\
		& = \sum_{m=0}^{k} \binom{N}{k} \binom{k}{m} (-1)^{k-m} \int d^{2} \beta P_{\rho} ( \beta ) e^{-\frac{(N-m) \eta}{N} |\beta-\alpha|^{2}}.
	\end{align}
Employing the convolution relation in Eq. (2) of main text and $\int d^{2} \beta P_{\rho} ( \beta ) = 1$, we obtain
	\begin{align}
		p_{k \neq N} [ \hat{D}^{\dag} ( \alpha ) \rho \hat{D} ( \alpha ) ] & = \sum_{m=0}^{k} T_{km} W_{\rho} ( \alpha; s_{m} ), \nonumber \\
		p_{N} [ \hat{D}^{\dag} ( \alpha ) \rho \hat{D} ( \alpha ) ] & = 1 + \sum_{m=0}^{N-1} T_{Nm} W_{\rho} ( \alpha; s_{m} ),
	\end{align}
where $s_{m} = 1 - \frac{2N}{(N-m)\eta}$ and $T_{km}$ is defined as
	\begin{align}
		T_{km} =
		\begin{cases}
			0 & \mbox{for $m>k$,} \\
			\binom{N}{k} \binom{k}{m} \frac{(-1)^{k-m} N \pi}{(N-m) \eta} & \mbox{for $m \leq k < N$,} \\
			1 & \mbox{for $m = k = N$.}
		\end{cases}
	\end{align}
We thus see that the counting statistics $p_{k}$ for a displaced state $\rho_\alpha\equiv\hat{D}^{\dag} ( \alpha ) \rho \hat{D} ( \alpha )$ is composed of quasiprobability functions via 
a triangular matrix $T$. The relation can be represented as 
	\begin{equation}
		\begin{pmatrix} p_{0} \\ p_{1} \\ \cdots \\ p_{N-1} \\ p_{N} \end{pmatrix} = T \begin{pmatrix} W_{\rho} ( \alpha; s_{0} ) \\ W_{\rho} ( \alpha; s_{1} ) \\ \cdots \\ W_{\rho} ( \alpha; s_{N-1} ) \\ 1 \end{pmatrix},
	\end{equation}
which leads to
	\begin{equation} \label{eq:sampling1}
		\begin{pmatrix} W_{\rho} ( \alpha; s_{0} ) \\ W_{\rho} ( \alpha; s_{1} ) \\ \cdots \\ W_{\rho} ( \alpha; s_{N-1} ) \\ 1 \end{pmatrix} = T^{-1} \begin{pmatrix} p_{0} \\ p_{1} \\ \cdots \\ p_{N-1} \\ p_{N} \end{pmatrix}.
	\end{equation}
using the inverse matrix $T^{-1}$. A triangular matrix is invertible if and only if all elements on its principal diagonal are non-zero \cite{SI-MatrixAnalysis}. We find that $T_{ii} \neq 0$ for all $i$, which means that we can always obtain $N$ different $s$-parametrized quasiprobability distributions from the photocounting statistics via $N$ on-off detectors using Eq.~\eqref{eq:sampling1}.

\section*{S7. Power of criteria using marginal distributions}

{\bf Marginal test}: Choosing $\beta_{i} = q_{i} + ip$ with the same $p$ for all $i$ in Eq. (3) of main text and integrating over $p$, we have
	$\int_{-\infty}^{\infty} dp \mathcal{M}_{ij}^{(n)} = \frac{\pi}{2} M_{\rho} ( \frac{q_{i}+q_{j}}{2}) e^{ - \frac{1}{2} |q_{i}-q_{j}|^{2}}.$
Under a coarse-graining with binning size $\sigma$, we choose test points as $q_{i} = (2m_{i}+k) \sigma + \delta$ with integer $m_{i}$, $k=0$ or 1, and $\delta\in[-\frac{\sigma}{2},\frac{\sigma}{2}]$. 
Integrating over $\delta$, we have $\int_{-\sigma/2}^{\sigma/2} d\delta \int_{-\infty}^{\infty} dp \mathcal{M}_{ij}^{(n)}= \frac{\pi}{2} M_{\rho}^{\sigma} [ m_{i} + m_{j} + k ] e^{- 2 (m_{i}-m_{j})^{2} \sigma^{2}}$,
where $M_{\rho}^{\sigma} [ n ] \equiv\int_{-\sigma/2}^{\sigma/2} d \delta M_{\rho} ( n \sigma + \delta )$ is the coarse-grained marginal distribution with $n$ the bin number. 
Our whole procedures corresponds to the classicality condition as $\sum_{i,j=1}^{n} c_{i}^{*} c_{j} \int d\delta\int dp\mathcal{M}_{ij}^{(n)} \geq 0$ thus constituting a matrix test with elements 
\begin{equation}
\mathcal{M}_{ij}^{(H)}\equiv\frac{\pi}{2} M_{\rho}^{\sigma} [ m_{i} + m_{j} + k ] e^{- 2 (m_{i}-m_{j})^{2} \sigma^{2}}.
\end{equation}

In the above, $k=0$ (1) corresponds to the case of choosing all even (odd)-numbered bins, making two different tests. If one chooses even-numbered ($n_1=2m_1$) and odd-numbered ($n_2=2m_2+1$) bins together, the middle bin ($\frac{n_1+n_2}{2}$) is not well-defined. 
Therefore, the matrix test in Eq. (S30) has been established according to the quadrature values $q_{i} = (2m_{i}+k) \sigma + \delta$ for a fixed $k$ (0 or 1), while $m_i$ may vary, from the beginning.

\subsubsection{Marginal distribution of Gaussian state}
Without loss of generality, we again deal with the case of a $x$-squeezed thermal state. Its marginal distribution is given by
	\begin{equation}
		M_{\sigma} ( q ) = \sqrt{\frac{2}{\pi}} e^{r-r_{c}} e^{-2e^{2(r-r_{c})}q^{2}},
	\end{equation}
with $\mu$ its purity, $r$ the squeezing strength and $r_{c} = - \frac{1}{2} \log \mu$ the critical squeezing strength for nonclassicality.  Considering the ratio $\mathcal{R} = \frac{M_{\rho} (-d) M_{\rho} (d)}{M_{\rho}^2 (0) \exp (-4d^{2} )}$, we find 
	\begin{equation}
		\mathcal{R} = e^{-4d^{2} [ e^{2(r-r_{c})} - 1  ]},
	\end{equation}
which indicates that all nonclassical Gaussian states, $r>r_c$, can be detected ($\mathcal{R}<1$) with an arbitrary non-zero $d$. 
More practically, we now examine a coarse-grained marginal distribution of Gaussian states.

\subsubsection{Coarse-grained marginal distribution of Gaussian state}
The coarse-grained marginal distribution of the Gaussian state is given by
	\begin{align}
		M_{\rho}^{\sigma} [n] = & \frac{1}{2} \mathrm{erf} [ \sqrt{2} e^{r-r_{c}} ( n + \frac{1}{2} ) \sigma ] \nonumber \\
		& - \frac{1}{2} \mathrm{erf} [ \sqrt{2} e^{r-r_{c}} ( n - \frac{1}{2} ) \sigma ].
	\end{align}
Here, $\mathrm{erf} (z) = \frac{2}{\sqrt{\pi}} \int_{0}^{z} e^{-t^{2}} dt$ is the error function that can be expanded as
	\begin{equation}
		\mathrm{erf} (z) = \frac{e^{-z^{2}}}{z \sqrt{\pi}} \sum_{n=0}^{\infty} (-1)^{n} \frac{(2n-1)!!}{(2z^{2})^{n}},
	\end{equation}
which indicates that $\mathrm{erf} (z) = \frac{e^{-z^{2}}}{z \sqrt{\pi}} \{ 1 + O(z^{-2}) \}$ for $z \gg 1$. Setting $z_{1} = \sqrt{2} e^{r-r_{c}} ( n - \frac{1}{2} ) \sigma$ and $z_{2} = \sqrt{2} e^{r-r_{c}} ( n + \frac{1}{2} ) \sigma$, we have
	\begin{equation}
		\frac{\mathrm{erf} (z_{2})}{\mathrm{erf} (z_{1})} = \frac{2n-1}{2n+1} e^{-4 e^{2 (r-r_{c})} n \sigma^{2}} \{ 1 + O ( (n\sigma)^{-2} ) \}.
	\end{equation}
which yields $\frac{\mathrm{erf} (z_{2})}{\mathrm{erf} (z_{1})} \ll 1$ for $r > r_{c}$ and $|n| \gg 1$. 
If we investigate the ratio $\mathcal{R}[n]\equiv \frac{M_{\rho}^{\sigma} [-2n] M_{\rho}^{\sigma} [2n]}{M_{\rho}^{\sigma} [0]^{2} \exp ( - 16n^{2}\sigma^{2} )}$ for  $r > r_{c}$ and $n\gg 1$, we obtain
	\begin{equation}
		\mathcal{R} \sim \bigg( \frac{e^{-z^{2}}}{2 z \sqrt{\pi}} \bigg)^{2} e^{16n^{2}\sigma^{2}},
	\end{equation}
with $z = \sqrt{2} e^{r-r_{c}} (2n-\frac{1}{2}) \sigma$.
For a sufficiently large $n$, we thus see $ \mathcal{R}\rightarrow 0$ for $r > r_{c}$. 
That is, $\mathcal{R}[n] <1$ in a certain range of large $n$.  
It indicates that all nonclassical Gaussian states can be witnessed by using coarse-grained marginal distribution with an arbitrary binning size $\sigma$.

\subsubsection{Marginal distribution of FSTS}
The marginal distribution of an arbitrary FSTS, $\rho = \sum_{j=0}^{N} \sum_{k=0}^{N} \rho_{jk} \ketbra{j}{k}$, is given by
	\begin{equation}
		M_{\rho} ( q ) = \sum_{j=0}^{N} \sum_{k=0}^{N} \rho_{jk} M_{\ketbra{j}{k}} ( q ),
	\end{equation}
where the marginal distribution for the operator $\ketbra{j}{k}$ is given by
	\begin{align}\label{eq:FDSm}
		M_{\ketbra{j}{k}} ( q ) & = \braket{q}{j} \braket{k}{q} \nonumber \\
		& = \sqrt{\frac{2}{\pi}} e^{-2q^{2}} \frac{H_{j} ( \sqrt{2} q )}{\sqrt{2^{j} j!}} \frac{H_{k} ( \sqrt{2} q )}{\sqrt{2^{k} k!}}.
	\end{align}
with the Hermite polynomial $H_{n} (x) = n! \sum_{m=0}^{\lfloor \frac{n}{2} \rfloor} \frac{(-1)^{m}}{m! (n-2m)!} (2x)^{n-2m}$ of degree $n$ \cite{SI-BarnettRadmore}. 

Using $H_{n} (x) = (2x)^{n} \{ 1 + O ( x^{-2} ) \}$ for $x \gg 1$, we have
	\begin{equation}
		M_{\ketbra{j}{k}} (q) = \sqrt{\frac{2}{\pi}} e^{-2q^{2}} \frac{(2q)^{j+k}}{\sqrt{j!k!}} \{ 1 + O ( q^{-2} ) \},
	\end{equation}
which yields
	\begin{equation} \label{eq:MDFDSL}
		M_{\rho} (q) = \rho_{NN} \sqrt{\frac{2}{\pi}} e^{-2q^{2}} \frac{(2q)^{2N}}{N!} \{ 1 + O ( q^{-1} ) \}.
	\end{equation}	
Looking into the ratio $\mathcal{R} = \frac{M_{\rho} ( 0 ) M_{\rho} ( 2d )}{M_{\rho}^2 ( d ) \exp ( - 4d^{2} )}$, we obtain $\mathcal{R} \propto d^{-2N}$ for a large $d \gg 1$.
Thus, $\mathcal{R}\rightarrow0$ for a very large $d$. This means that the nonclassicality of every FSTS can be verified, $\mathcal{R}<1$, in a certain range of large $d$. 
More practically, we now examine a coarse-grained marginal distribution of FSTS.

\subsubsection{Coarse-grained marginal distribution of FSTS}
The coarse-grained marginal distribution of FSTS is given by
	\begin{equation} \label{eq:CGMDFDS}
		M_{\rho}^{\sigma} [ n ] = \sum_{j=0}^{N} \sum_{k=0}^{N} \rho_{jk} \int_{-\frac{\sigma}{2}}^{\frac{\sigma}{2}} d\delta M_{\ketbra{j}{k}}  ( n\sigma + \delta ),
	\end{equation}
using $M_{\ketbra{j}{k}}  (q)$ in Eq.~\eqref{eq:FDSm}.
We are going to deal with $M_{\rho}^{\sigma} [ n ] $ for a large $n$, so with Eq.~\eqref{eq:MDFDSL} in mind, we first look at the integration
	\begin{align}
		& \int_{x_{1}}^{x_{2}} dx e^{-2x^{2}} x^{m} \nonumber \\
		& = 2^{-\frac{m+3}{2}} \{ \Gamma ( \frac{m+1}{2}, 2x_{1}^{2} ) - \Gamma ( \frac{m+1}{2}, 2x_{2}^{2} ) \}.
	\end{align}
Here $\Gamma ( a, x ) = \int_{x}^{\infty} e^{-t} t^{a-1} dt$ is an incomplete Gamma function \cite{SI-Arfken}, which can be expanded as
	\begin{equation} \label{eq:IGF}
		\Gamma ( a, x ) = x^{a-1} e^{-x} \sum_{n=0}^{\infty} \frac{(a-1)!}{(a-1-n)!} x^{-n}.
	\end{equation}
Eq.~\eqref{eq:IGF} indicates that $\Gamma ( a, x ) = x^{a-1} e^{-x} \{ 1 + O ( x^{-1} ) \}$ for $x \gg 1$. Putting $x_{1} = (n-\frac{1}{2}) \sigma$ and $x_{2} = (n+\frac{1}{2}) \sigma$, we thus have
	\begin{equation}
		\frac{\Gamma ( \frac{m+1}{2}, 2x_{2}^{2} )}{\Gamma ( \frac{m+1}{2}, 2x_{1}^{2} )} = e^{- 4n\sigma^{2}} \{ 1 + O ( n^{-1} ) \},
	\end{equation}
which yields
	\begin{align} \label{eq:IFL}
		& \int_{(n-\frac{1}{2}) \sigma}^{(n+\frac{1}{2}) \sigma} dx e^{-2x^{2}} x^{m} \nonumber \\
		& \approx 2^{-\frac{m+3}{2}} \exp\bigg(- \frac{(2n-1)^{2}\sigma^{2}}{2}\bigg) \bigg(\frac{(2n-1)^{2}\sigma^{2}}{2}\bigg)^{\frac{m-1}{2}},
	\end{align}
for $n \gg 1$. Using Eqs.~\eqref{eq:MDFDSL}, \eqref{eq:CGMDFDS} and \eqref{eq:IFL}, we thus obtain 
	\begin{align}
		M_{\rho}^{\sigma} [ n ] & \approx \rho_{NN} \sqrt{\frac{2}{\pi}} \frac{1}{N!} 2^{\frac{2N-3}{2}} \nonumber \\
		& \times \exp \bigg( - \frac{(2n-1)^{2}\sigma^{2}}{2} \bigg) \bigg( \frac{(2n-1)^{2}\sigma^{2}}{2} \bigg)^{\frac{2N-1}{2}},
	\end{align}
for $n \gg 1$. Considering the ratio $\frac{M_{\rho}^{\sigma} [ 0 ] M_{\rho}^{\sigma} [ 2n ]}{M_{\rho}^{\sigma} [ n ]^{2} \exp ( - 4n^{2}\sigma^{2} )}$, we thus see $\mathcal{R} \propto ( n \sigma )^{1-2N}$ for $n \gg 1$, which indicates that the nonclassicality of every FSTS can be verified by using the coarse-grained marginal distribution with an arbitrary finite binning size $\sigma$.

\section*{S8. Error analysis and practical examples}
The data acquisition procedure can be thought of as picking up outcomes randomly from a multinomial distribution. Let us assume that the total number of possible outcomes is $k$ and the probability to obtain $i$-th outcome is $p_{i}$ with $\sum_{i=1}^{k} p_{i} = 1$. If we pick up outcomes $N_s$ times and the number of observation for $i$-th outcome is $X_{i}$, the variance of the probability estimator $\frac{X_{i}}{N_s}$ and the covariance between the probability estimators $\frac{X_{i}}{N_s}$ and $\frac{X_{j}}{N_s}$ are given by $\frac{p_{i}(1-p_{i})}{N_s}$ and $- \frac{p_{i}p_{j}}{N_s}$, respectively \cite{SI-Forbes}. For the test based on the on-off detectors, we obtain $s$-parametrized quasiprobability distributions by a linear transformation of the measured click-counting probability as described in Sec. S5. For a linear combination $Z = \sum_{i} a_{i} X_{i}$, the variance of $Z$ is given by $( \Delta Z )^{2} = \sum_{i} a_{i}^{2} ( \Delta X_{i} )^{2} + \sum_{i} \sum_{i \neq j} a_{i} a_{j} \mathrm{Cov} ( X_{i}, X_{j} )$ where $\Delta x$ and $\mathrm{Cov} (x,y)$ represent the standard deviation of $x$ and the covariance between $x$ and $y$, respectively. Using the propagation of the uncertainty, we can estimate the statistical uncertainty of the obtained $s$-parametrized quasiprobability distibutions.

Furthermore, the statistical uncertainty of the ratio $R = \frac{AB}{C^{2}}$ can be estimated as
	\begin{align}
		\delta R & = \frac{(A + \delta A)(B + \delta B)}{(C + \delta C)^{2}} - \frac{AB}{C^{2}} \nonumber \\
		& \approx \frac{AB}{C^{2}} \bigg( \frac{\delta A}{A} + \frac{\delta B}{B} - \frac{2 \delta C}{C} \bigg),
	\end{align}
which yields
	\begin{align}
		\bigg( \frac{\Delta R}{R} \bigg)^{2} & \approx \bigg( \frac{\Delta A}{A} \bigg)^{2} + \bigg( \frac{\Delta B}{B} \bigg)^{2} + \bigg( \frac{2 \Delta C}{C} \bigg)^{2} \nonumber \\
		& + \frac{2 \mathrm{Cov} (A,B)}{AB} - \frac{4 \mathrm{Cov} (B,C)}{BC} - \frac{4 \mathrm{Cov} (C,A)}{CA}.
	\end{align}
We can also estimate the statistical uncertainty of the ratio for the test based on the coarse-grained homodyne detection by a similar error analysis.

We now illustrate the power of two practical tests using the on-off detectors and the homodyne detection, respectively, including the analysis of error due to a finite data acquisition. 

{\bf On-off detector test}---
We first show the results for the case of Fock state $|n\rangle$ under a 50\% loss channel, which makes all Wigner functions positive definite. 
In Fig. S3, we demonstrate the detection of nonclassicality using $N=2$ on-off detectors with a data number $N_s\sim10^5$. The red curve represents $\mathcal{R} = \frac{W_{\rho} ( 0 ) W_{\rho} ( 2d )}{W_{\rho} ( d )^{2} \exp ( - 4d^{2} )}$ as a function of displacement $d$, with the grey shades representing the error size $\Delta \mathcal{R}$ for each $d$. We see that there always exists a range of $d$ for each state in which the value $\mathcal{R}$ is well below 1 with the error $\Delta \mathcal{R}$ included, that is, $1>\mathcal{R}+\Delta \mathcal{R}$. For instance, we have  $\frac{1-\mathcal{R}}{\Delta \mathcal{R}}=2.21$ at $d=1.87$ for the noisy Fock state $|3\rangle$, so the signal beats the error over 2 standard deviation. Interestingly, the signal to noise ratio increases with $n$, as seen from figures, as $\frac{1-R}{\Delta \mathcal{R}}$=3.7, 5.61, and 7.94 at $d$=1.89, 1.91, and 1.94, respectively, for the states $|4\rangle,  |5\rangle$, and $|6\rangle$.
	\begin{figure}[ht]
          \includegraphics[scale=0.4]{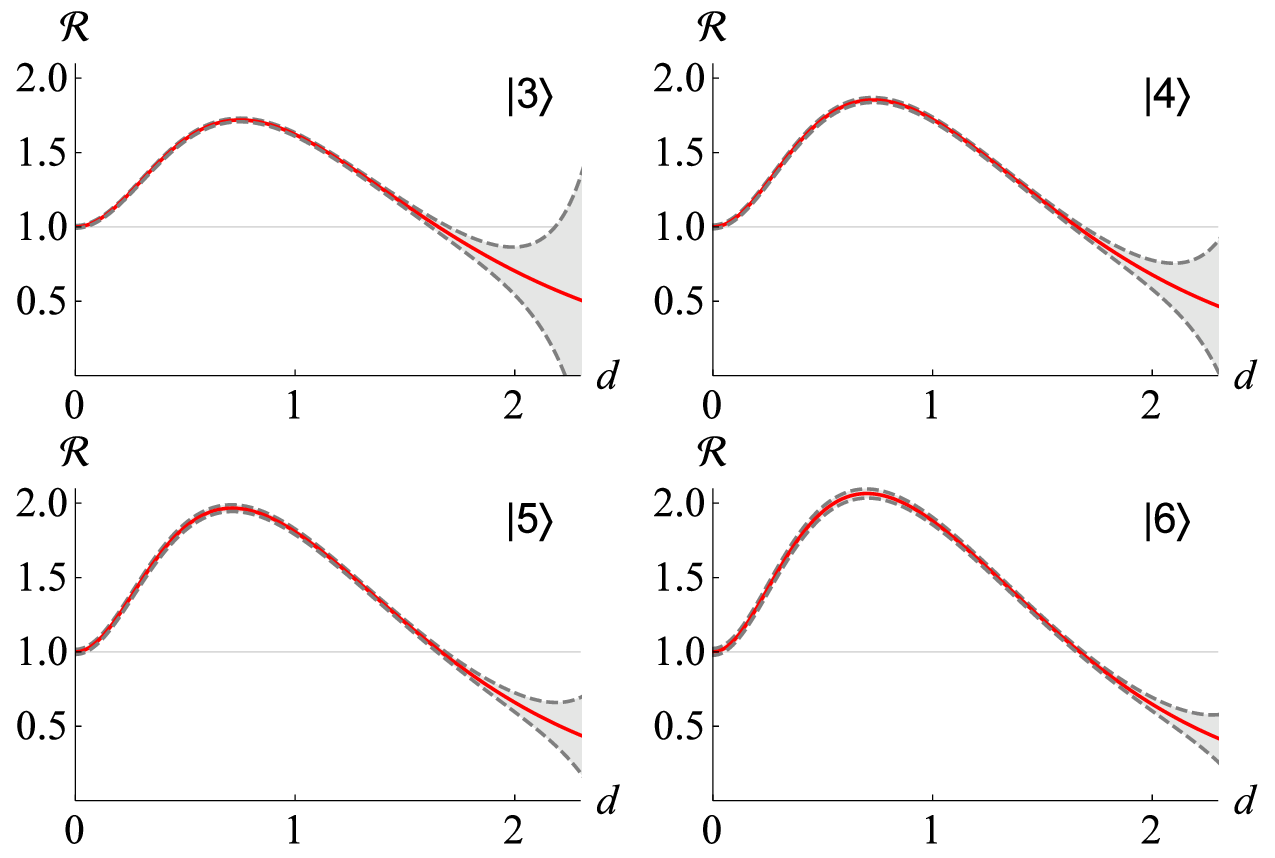}
		\caption{Testing a Fock state $|n\rangle$ under a 50\% loss channel using two on-off detectors with detector efficiency $\eta=0.9$ \cite{SI-Nam}. 
Each red curve represents $\mathcal{R} = \frac{W_{\rho} ( 0 ) W_{\rho} ( 2d )}{W_{\rho} ( d )^{2} \exp ( - 4d^{2} )}$ as a function of displacement $d$. 
Grey shades represent the size of error due to finite data $\sim10^5$. $\mathcal{R}<1$ confirms nonclassicality.}
		\label{fig:DALF}
	\end{figure}

We further illustrate the on-off test for the case of Fock states under a 50\% loss channel now mixed with a thermal photon $\bar{n}=0.05$. Previously we have chosen three displacements as $\{0,d,2d\}$. The test can be enhanced by introducing one more parameter $d_1$ for optimization choosing points as $\{d_1,d_1+d,d_1+2d\}$. Thus $d_1$ represents the starting point whereas $d$ is the distance among them. Adopting this strategy, we analyze the case of lower detector efficiency $\eta=0.75$ in Fig. S4.  We again confirm the successful detection of nonclassicality reliably against noise with a lower detector efficiency for a broad range of displacements.

	\begin{figure}[ht]
          \includegraphics[scale=0.4]{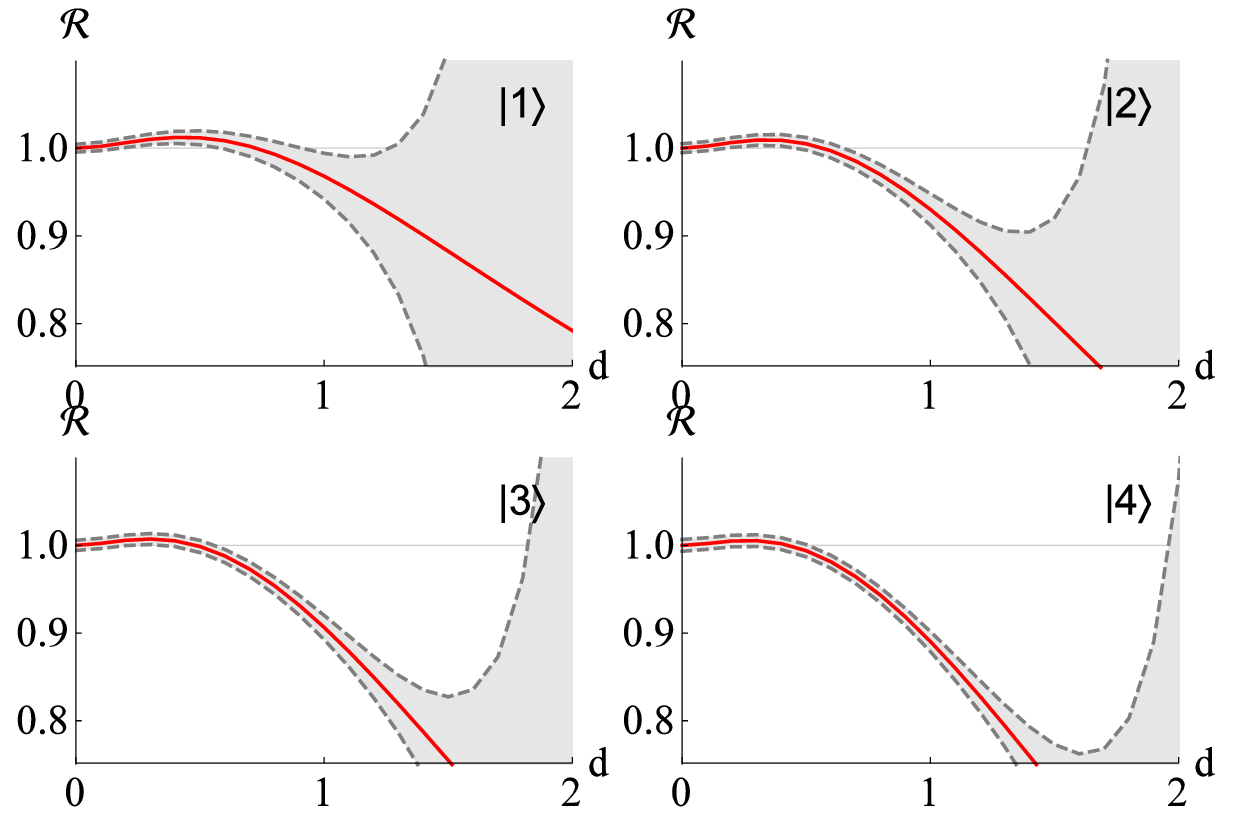}
		\caption{Testing a Fock state $|n\rangle$ under a 50\% loss channel mixed with a thermal photon $\bar{n}=0.05$ using two on-off detectors with detector efficiency $\eta=0.75$. 
Each red curve represents $\mathcal{R} = \frac{W_{\rho} ( d_1 ) W_{\rho} ( d_1+2d )}{W_{\rho} (d_1+d )^{2} \exp ( - 4d^{2} )}$ as a function of displacement $d$ with $d_1=1$. 
Grey shades represent the size of error due to finite data $\sim10^5$. $\mathcal{R}<1$ confirms nonclassicality.}
		\label{fig:75}
	\end{figure}

{\bf Homodyne test}---
We now demonstrate the detection of nonclassicality using a coarse-grained marginal distribution with a binning size $\sigma=0.1$ and a data number $N_s\sim10^6$. In Fig. S5, we show the results for the same noisy Fock states as in Fig. S3. 
Black circles represent $\mathcal{R} = \frac{M_{\rho}^{\sigma} [ 2m_1 ] M_{\rho}^{\sigma} [ 2m ]}{M_{\rho}^{\sigma} [ m_{1}+m ]^{2} \exp ( - 4(m_{1}-m)^{2}\sigma^{2} )}$ at each value of $m$ with $m_1=10$, which corresponds to the choice of three quadratures $q_1=2m_1\sigma,q_2=(m_1+m)\sigma,$ and $q_3=2m\sigma$ for test. The grey shades represent the error size $\Delta \mathcal{R}$ due to finite data $N_s\sim10^6$. We again see that there always exists a range of $m$ for each state to achieve $1>\mathcal{R}+\Delta \mathcal{R}$. For instance, we have  $\frac{1-R}{\Delta \mathcal{R}}=21.34$ at $m=4$ for the noisy Fock state $|3\rangle$, so the signal beats the error over 21 standard deviation. For other Fock states $|4\rangle,  |5\rangle$, and $|6\rangle$, the signal to noise ratio is $\frac{1-R}{\Delta \mathcal{R}}$=17.73, 11.84, and 6.48 at $m$=5, 6, and 7, respectively.

	\begin{figure}[ht]
          \includegraphics[scale=0.4]{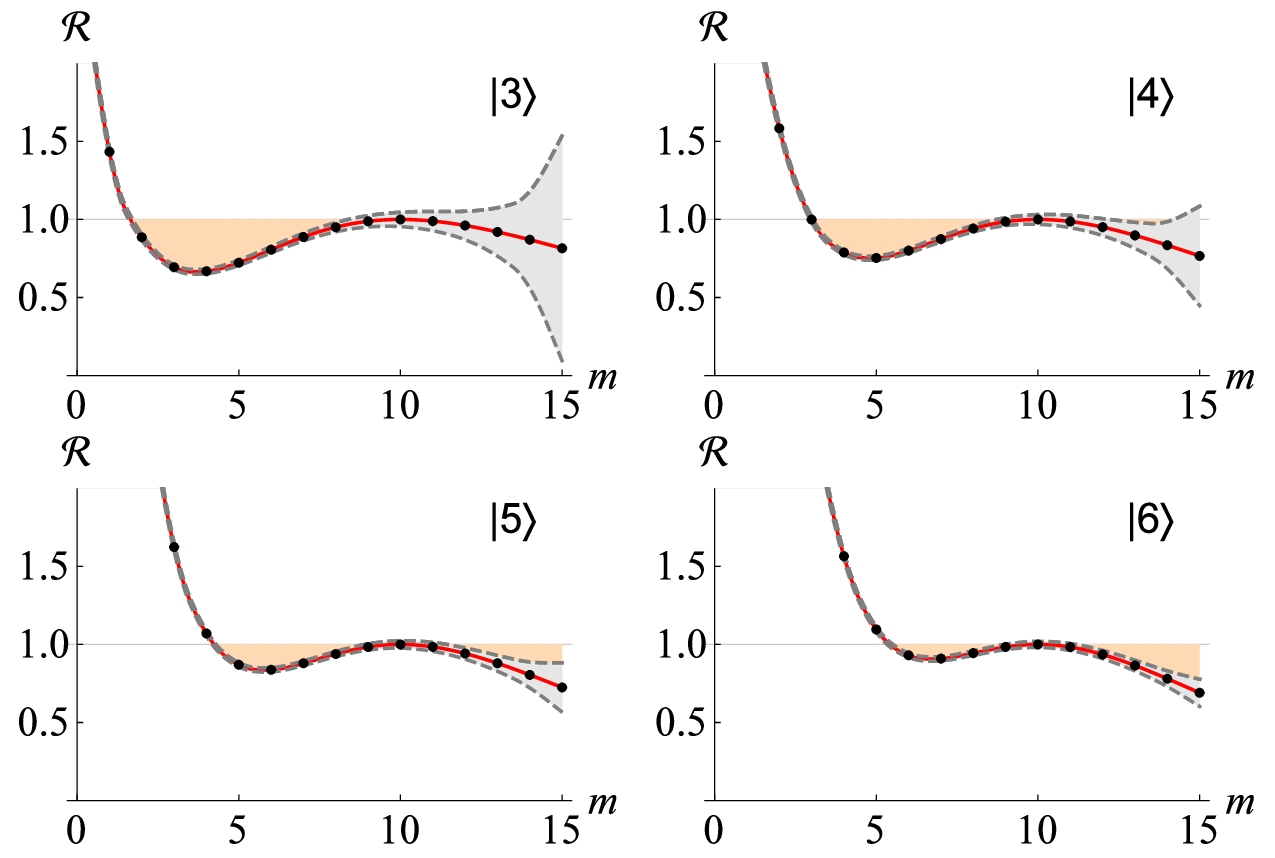}
		\caption{Testing a Fock state $|n\rangle$ under a 50\% loss channel using homodyne detection with a binning size $\sigma = 0.1$. 
Black dots represent $\mathcal{R} = \frac{M_{\rho}^{\sigma} [ 2m_{1} ] M_{\rho}^{\sigma} [ 2m ]}{M_{\rho}^{\sigma} [m_{1}+m ]^{2} \exp ( - 4(m_{1}-m)^{2}\sigma^{2} )}$ against $m$ with $m_{1}=10$.  Red curves join the black dots for visual guide. $m$ designates a bin number corresponding to quadrature $q=m\sigma$. Grey shades represent the size of error due to finite data $\sim10^6$. $\mathcal{R}<1$ confirms nonclassicality. 
The results clearly manifest nonclassicality beating error in the orange region, where the error bar is negligible so not appreciable in the plot.}
		\label{fig:HDLF}
	\end{figure}

	\begin{figure}[ht]
          \includegraphics[scale=0.4]{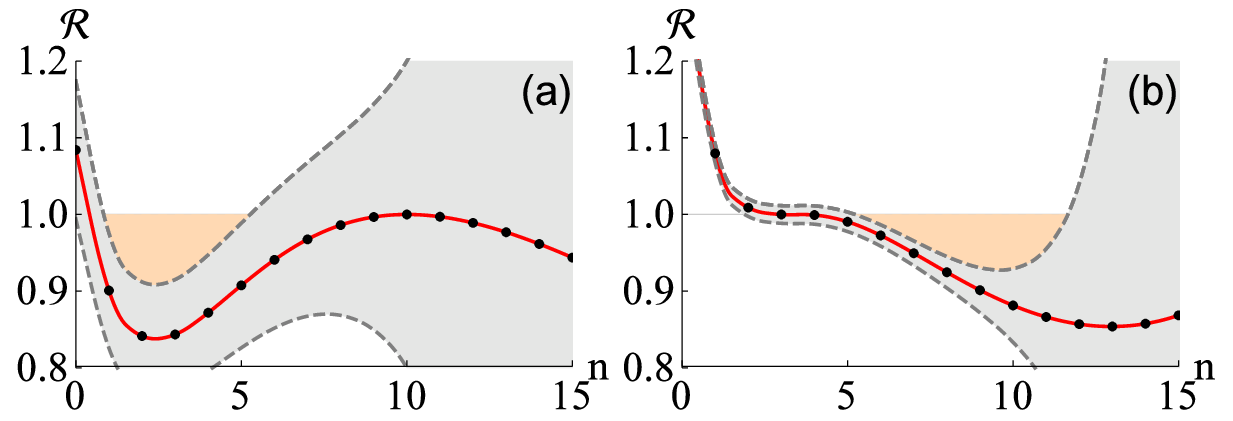}
		\caption{Testing (a) a noisy Fock state $f|1\rangle\langle1|+(1-f)|0\rangle\langle0|$ at $f=0.3$ and (b) a single photon state under a 50\% loss channel mixed with a thermal photon ${\bar n}=0.1$ using homodyne detection with a binning size $\sigma = 0.1$.  
Black dots represent $\mathcal{R} = \frac{M_{\rho}^{\sigma} [ 2n_{1} ] M_{\rho}^{\sigma} [ 2n ]}{M_{\rho}^{\sigma} [n_{1}+n ]^{2} \exp ( - 4(n_{1}-n)^{2}\sigma^{2} )}$ against $n$ with (a) $n_{1}=10$ and (b) $n_{1}=3$.  Red curves join the black dots for visual guide. $n$ designates a bin number corresponding to quadrature $q=n\sigma$. Grey shades represent the size of error due to finite data $\sim10^6$. $\mathcal{R}<1$ confirms nonclassicality.
The results clearly manifest nonclassicality beating error in the orange region.}
		\label{fig:CGHDNF}
	\end{figure}

As seen from the above examples, the corase-grained homodyne test is very robust against experimental imperfections. We further demonstrate its practical power with other examples. 
Fig. S6 (a) confirms the case of a noisy single-photon state $f|1\rangle\langle1|+(1-f)|0\rangle\langle0|$ at $f=0.3$, and Fig. S6 (b) for a single photon state under a 50\% loss channel mixed with a thermal photon ${\bar n}=0.1$. 
For comparison, there have been some protocols proposed to distill squeezing by which one may confirm nonclassicality upon the postselected (distilled) copies. For example, R. Filip remarkably proposed a distillation of squeezing for a noisy single-photon state \cite{SI-Filip2014}. 
Its single-copy distillation (Fig. 1 of \cite{SI-Filip2014}) entails a low probability of success, e.g. $<10^{-2}$ for the case of $f=0.3$, with distilled squeezing low to detect. 
In constrast, our method does not require postselection of data and manifests nonclassicality robust against practical errors.

	\begin{figure}[ht]
          \includegraphics[scale=0.4]{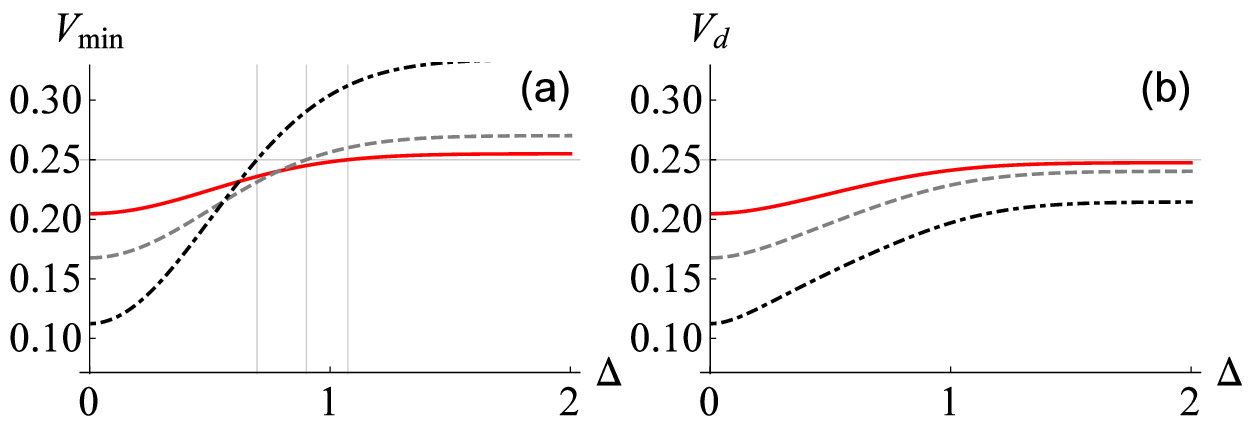}
		\caption{(a) Minimum quadrature variance $V_{\mathrm{min}}$ of a phase-diffused squeezed vacuum with squeezing parameter $r = 0.1$ (red solid), $r = 0.2$ (gray dashed) and $r = 0.4$ (black dot-dashed) against the size (standard deviation) of phase diffusion $\Delta$. The squeezing disappears at $\Delta > 1.074$, $\Delta > 0.901$ and $\Delta >0.696$ for squeezed states with squeezing parameter $r = 0.1$, $r = 0.2$ and $r =0.4$, respectively. 
(b) Distilled variance $V_d$ for each phase-diffused state in (a) applying the protocol in \cite{SI-Filip2013}.}
		\label{fig:PDSV1}
	\end{figure}

	\begin{figure}[ht]
          \includegraphics[scale=0.4]{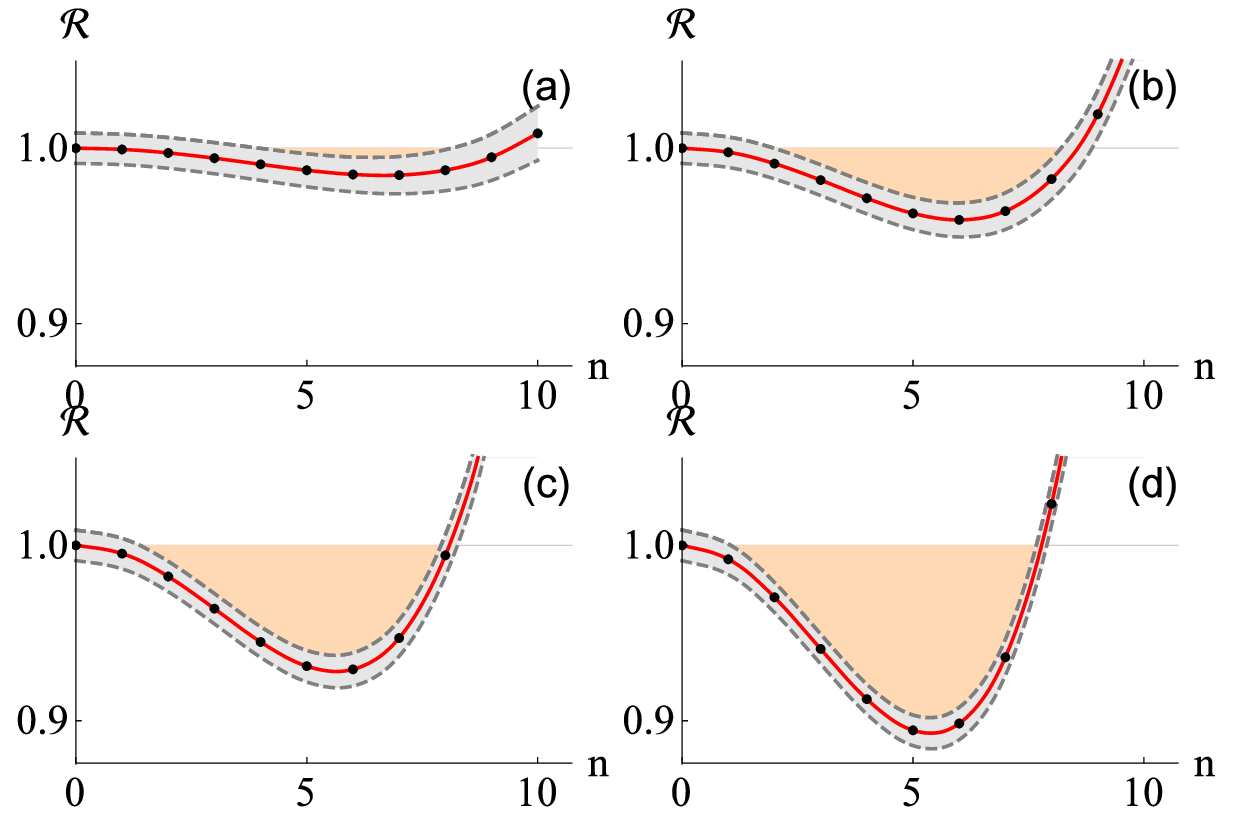}
		\caption{Homodyne test for a phase-diffused squeezed state at phase-diffusion $\Delta=1.2$, with (a) $r=0.1$ (b) $r=0.2$ (c) $r=0.3$ (d) $r=0.4$. $\mathcal{R} = \frac{M_{\rho}^{\sigma} [ -n ] M_{\rho}^{\sigma} [ n ]}{M_{\rho}^{\sigma} [0]^{2} \exp ( - 4n^{2}\sigma^{2} )}$ (red solid) against $n$ with binning size $\sigma = 0.1$. $n$: bin number for quadrature $q=n\sigma$. Grey shades represent the size of error due to finite data $\sim10^6$. $\mathcal{R}<1$ confirms nonclassicality.}
		\label{fig:PDV}
	\end{figure}

Next we move on to consider another practical noise, i.e. phase-diffusion of an optical signal. 
A phase diffusion can be described as
	\begin{equation}
		\mathcal{D} [ \rho ] = \int d \phi \sqrt{\frac{1}{2 \pi \Delta^{2}}} \exp \bigg( - \frac{\phi^{2}}{2 \Delta^{2}} \bigg) e^{i \hat{n} \phi} \rho e^{-i \hat{n} \phi},
	\end{equation}
where $\Delta$ is the standard deviation of random phase-shift. The phase diffusion affects the marginal distribution of a quantum state $\rho$ as
	\begin{equation}
		M_{\mathcal{D}[\rho]} ( x_{\theta} ) = \int d \phi \sqrt{\frac{1}{2 \pi \Delta^{2}}} \exp \bigg( - \frac{\phi^{2}}{2 \Delta^{2}} \bigg) M_{\rho} ( x_{\theta + \phi} ),
	\end{equation}
where $M_{\rho} ( x_{\theta} )$ represents the marginal distribution of the quantum state $\rho$ for the quadrature ${\hat x}_{\theta}=\frac{\hat{a}e^{-i\theta}+\hat{a}^\dag e^{i\theta}}{2}$.

As shown in Fig. S7 (a), the phase-diffusion destroys squeezing at $\Delta > 1.074$, $\Delta > 0.901$ and $\Delta >0.696$ for squeezed states with squeezing parameter $r = 0.1$, $r = 0.2$ and $r =0.4$, respectively. 
The squeezing can be distilled after phase-diffusion, as shown in Fig. S7 (b), e.g. by applying the protocol in \cite{SI-Filip2013} that generally requires multi copies of the same nonclassical states and postselection. 
Our homodyne test directly confirms nonclassicality for these phase-diffused squeezed states. Fig. S8 shows a strong result from our test for the case of $\Delta=1.2$, at which all considered states completely lose squeezing, against coarse-graining of homodyne data and finite data acqusition.

{\it Comparison with moment test}---The above examples illustrate that our homodyne test verifies nonclassicality when the usual squeezing test by homodyne detection fails. 
We may further compare our test with moment-based homodyne tests. Suppose that one obtains a marginal distribution $M(q)=\int dp W_{\rho} (q,p)$ from a homodyne measurement, where $W_{\rho} (q,p)$ is the Wigner function.  
The distribution $M(q)$ can be related to the marginal distribution ${\tilde M}({\tilde q})=\int dp P_{\rho} (q,p)$ of the Sudarshan-Glauber $P$-function $P_{\rho} (q,p)$. That is, $M(q)=\sqrt{\frac{2}{\pi}}\int d{\tilde q} {\tilde M}({\tilde q})e^{-2(q-{\tilde q})^2}$. 
If the state is classical, the distribution ${\tilde M}({\tilde q})$ must be positive-definite. This can be checked in terms of moments $\langle {\tilde q}^k\rangle\equiv \int d{\tilde q} {\tilde M}({\tilde q}) {\tilde q}^k$. We can readily see that an $n\times n$ matrix $\cal {Q}$, whose elements are given by ${\cal Q}_{ij}=\langle {\tilde q}^{i+j} \rangle$, must be positive definite. For example, the violation of the condition ${\cal Q}\ge 0$ at $n=2$ represents the usual quadrature squeezing.

	\begin{figure}
		\includegraphics[scale=0.4]{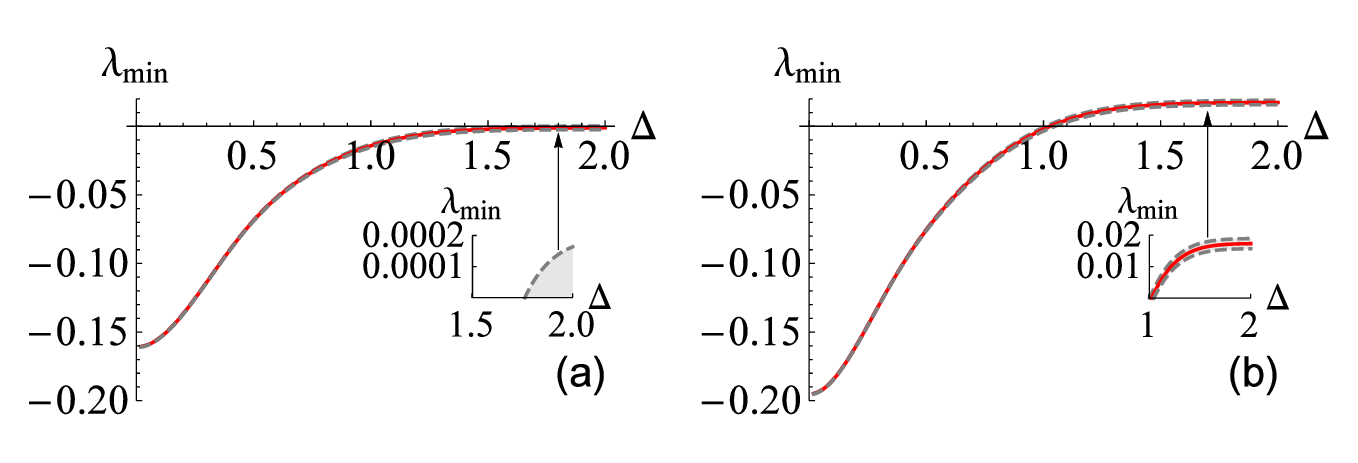}
		\caption{Minimum eigenvalue $\lambda_{\min}$ (red solid) of the $4\times4$ ${\cal Q}$-matrix for the phase-diffused squeezed states with squeezing (a) $r=0.4$ and (b) $r=0.5$ against the phase diffusion $\Delta$. A negative $\lambda_{\min}<0$ verifies nonclassicality. The black dashed curves show the results including the error due to a finite data $\sim10^6$ and binning size $\sigma = 0.1$. We see that the moment test fails at a sufficiently large phase-diffusion.}
	\end{figure}

	\begin{figure}[ht]
          \includegraphics[scale=0.45]{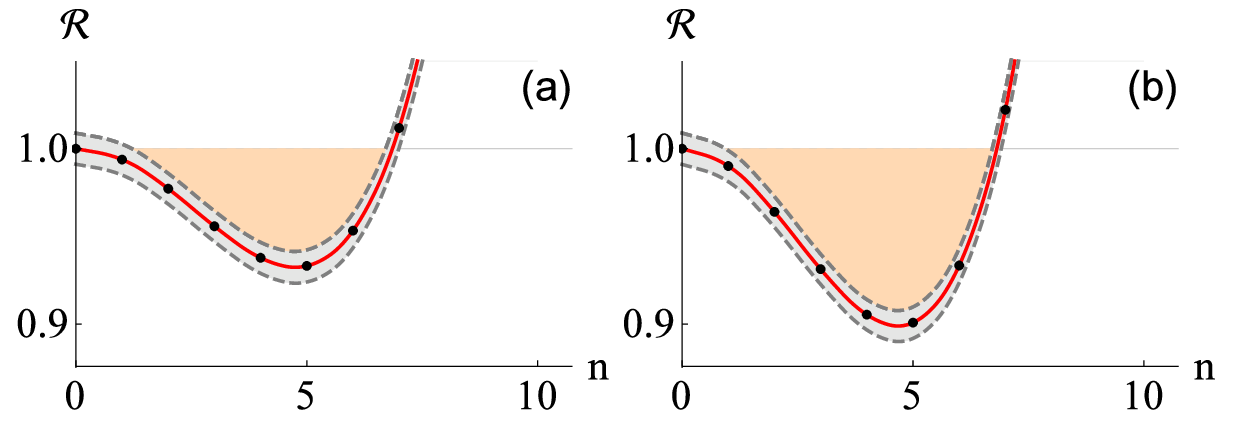}
		\caption{Homodyne test for a phase-diffused squeezed state at phase-diffusion $\Delta=2.0$, with (a) $r=0.4$ and (b) $r=0.5$ $\mathcal{R} = \frac{M_{\rho}^{\sigma} [ -n ] M_{\rho}^{\sigma} [ n ]}{M_{\rho}^{\sigma} [0]^{2} \exp ( - 4n^{2}\sigma^{2} )}$ (red solid) against $n$ with binning size $\sigma = 0.1$. $n$: bin number for quadrature $q=n\sigma$. Grey shades represent the size of error due to finite data $\sim10^6$. $\mathcal{R}<1$ confirms nonclassicality.}
	\end{figure}
We have performed the analysis on the phase-diffused squeezed states using $\cal {Q}$ matrix test. This test fails to detect nonclassicality at $n=2$ (squeezing) and $n=3$ for a severely decohered state, so we move to the level $n=4$. 
Fig. S9 shows the results for the squeezed state with $r=0.4$ and $r=0.5$ against the phase diffusion $\Delta$. For a fair comparison, we include the effects of finite data $\sim10^6$ and the coarse-graining $\sigma=0.1$ for the error analysis.
To confirm nonclassicality, the least eigenvalue (red solid) of the matrix $\cal {Q}$ must be negative. Consdering the error level (black dashed), we see that the test fails at a large phase-diffusion, $\Delta>1.76$ ($r=0.4$ case) and $\Delta>1$ ($r=0.5$ case). 
One may try to enhance the test by further going up to the higher level of $n$, which however is not much favoarable in the presence of errors due to finite data and coarse-graining. In contrast, as we show in Fig. S10, our homodyne test clearly manifests nonclassicality even at the large phase-diffusion $\Delta=2$ for the same squeezed states.

\section*{S9. Estimating the nonclassical depth}
Our formulation also provides an estimate for another nonclassicality measure, i.e. nonclassical depth \cite{SI-Lee1991}, as follows. 

	\begin{figure}[ht]
         \includegraphics[scale=0.4]{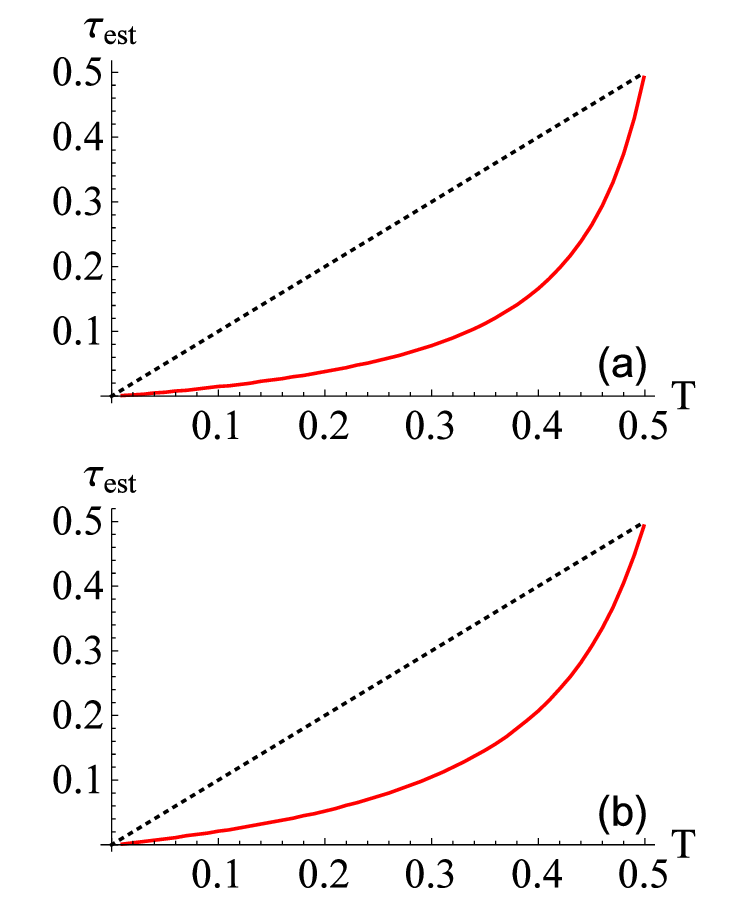}
		\caption{Estimated nonclassical depth (red solid) using Eq. (S51) by optimizing  $\alpha_{2}=d$ with $ \alpha_{1}=0$ for the Fock states  (a) $|1\rangle$ and (b) $|2\rangle$ under a loss channel with transmittance $T$. Black dotted curves represent the actual nonclassical depth.}
	\end{figure}

If the nonclassical depth of a quantum state $\rho$ is $\tau$, the $s$-parametrized quasi-probability function of the quantum state, i.e., $W_{\rho} ( \alpha ; s )$, has to satisfy
	\begin{equation} \label{eq:NDE}
		\frac{W_{\rho} ( \alpha_{1} ; s ) W_{\rho} ( \alpha_{2} ; s )}{W_{\rho} ( \frac{\alpha_{1}+\alpha_{2}}{2} )^{2}} \geq \exp \bigg( - \frac{|\alpha_{1}-\alpha_{2}|^{2}}{s_{\tau}-s} \bigg).
	\end{equation}
for $s \leq s_{\tau} \equiv 1 - 2 \tau$. Its derivation goes as follows. By the definition of nonclassical depth, the $s$-parametrized quasi-probability function becomes non-negative for $s \leq s_{\tau} \equiv 1 - 2 \tau$. 
Using the convolution between the $s$-parmetrized functions ($s \leq s'$)
	\begin{equation}
		W_{\rho} ( \alpha; s ) = \frac{2}{\pi ( s^{\prime} - s )} \int d^{2} \beta W_{\rho} ( \beta ; s^{\prime} ) \exp \bigg( - \frac{2 | \alpha - \beta |^{2}}{s^{\prime} - s} \bigg),
	\end{equation}
we obtain
	\begin{eqnarray}
		&&\frac{\pi ( s^{\prime} - s )}{2} \int d^{2} \beta W_{\rho} ( \beta ; s_{\tau} ) \bigg| \sum_{i} c_{i} \exp \bigg( - \frac{| \alpha_{i} - \beta |^{2}}{s_{\tau} - s} \bigg) \bigg|^{2} \nonumber\\&&= \sum_{i} \sum_{j} c_{i} c_{j}^{*} W_{\rho} \bigg( \frac{\alpha_{i}+\alpha_{j}}{2} ; s \bigg) \exp \bigg( - \frac{1}{s_{\tau} - s} \frac{|\alpha_{i}-\alpha_{j}|^{2}}{2} \bigg)\nonumber\\&& \geq 0,
	\end{eqnarray}
which yields Eq.~\eqref{eq:NDE}.

In Fig. S11, we illustrate the estimate of nonclassical depth using Eq. (S51) for Fock states (a) $|1\rangle$ and (b) $|2\rangle$ under a loss mechanism with transmittance $T$. We show the analysis for the case of positive Wigner function. The red solid curves represent our estimate based on Eq. (S51), whereas the black dotted curves represent the actual nonclassical depth.

\bibliographystyle{apsrev}

\end{document}